# Anomalous dynamics of the extremely compressed magnetosphere during 21 January 2005 magnetic storm

A.V. Dmitriev[1,2], A. V. Suvorova[1,2], J.-K. Chao[1], C. B. Wang[3], L. Rastaetter[4], M. I. Panasyuk[2], L. L. Lazutin[2], A. S. Kovtyukh[2], I. S. Veselovsky[2,5], I. N. Myagkova[2]

[1]*Institute of Space Science, National Central University, Chung-Li, Taiwan*

[2] *Lomonosov Moscow State University Skobeltsyn Institute of Nuclear Physics (MSU SINP), Moscow, Russia*

[3]*CAS Key Lab of Geospace Environment, Department of Geophysics and Planetary Science, University of Science and Technology of China, Hefei, China*

[4]*Goddard Space Flight Center, Maryland, USA*

[5]*Space Research Institute (IKI), Russian Academy of Sciences, Moscow, Russia*

Short title: EXTREME MAGNETOSPHERIC COMPRESSION

A. V. Dmitriev, Institute of Space Science, National Central University, Chung-Li, 320, Taiwan, also at Lomonosov Moscow State University Skobeltsyn Institute of Nuclear Physics (MSU SINP), Russia (e-mail: dalex@jupiter.ss.ncu.edu.tw)

A. V. Suvorova, Institute of Space Science, National Central University, Chung-Li, 320, Taiwan, also at Lomonosov Moscow State University Skobeltsyn Institute of Nuclear Physics (MSU SINP), Russia (e-mail: alla@jupiter.ss.ncu.edu.tw)

J.-K. Chao, Institute of Space Science, National Central University, Chung-Li, 320, Taiwan (e-mail: jkchao@jupiter.ss.ncu.edu.tw)

C. B. Wang, CAS Key Lab of Geospace Environment, Department of Geophysics and Planetary Science, University of Science and Technology of China, Hefei, China. (e-mail: cbwang@ustc.edu.cn)

L. Rastaetter, Goddard Space Flight Center, Maryland, USA (e-mail: Lutz.Rastaetter@nasa.gov)

M. I. Panasyuk, Lomonosov Moscow State University Skobeltsyn Institute of Nuclear Physics (MSU SINP), Russia (e-mail: panasyuk@srdmail.sinp.msu.ru)

L. L. Lazutin, Lomonosov Moscow State University Skobeltsyn Institute of Nuclear Physics (MSU SINP), Russia (e-mail: lll@srd.sinp.msu.ru)

A. S. Kovtyukh, Lomonosov Moscow State University Skobeltsyn Institute of Nuclear Physics (MSU SINP), Russia (e-mail: kovtyukhas@mail.ru)

I. S. Veselovsky, Lomonosov Moscow State University Skobeltsyn Institute of Nuclear Physics (MSU SINP), also at Space Research Institute (IKI), Russian Academy of Sciences, Moscow, Russia (e-mail: veselov@dec1.sinp.msu.ru)

I. N. Myagkova, Lomonosov Moscow State University Skobeltsyn Institute of Nuclear Physics (MSU SINP), Russia (e-mail: irina@srd.sinp.msu.ru)

**Abstract**

Dynamics of the dayside magnetosphere and proton radiation belt was analyzed during unusual magnetic storm on 21 January 2005. We have found that during the storm from 1712 to 2400 UT, the subsolar magnetopause was continuously located inside geosynchronous orbit due to strong compression. The compression was found to be extremely strong from 1846 to 2035 UT when the dense plasma of fast erupting filament produced the solar wind dynamic pressure *Pd* peaked up to >100 nPa and, in the first time, the upstream solar wind was observed at geosynchronous orbit during almost 2 hours. Under the extreme compression, the outer magnetosphere at $L > 5$ was pushed inward and the outer radiation belt particles with energies of several tens of keV moved earthward, became adiabatically accelerated and accumulated in the inner magnetosphere at $L < 4$ that produced the intensified ring current with an exceptionally long lifetime. The observations were compared with predictions of various empirical and first principles models. All the models failed to predict the magnetospheric dynamics under the extreme compression when the minimal magnetopause distance was estimated to be ~3 Re. The inconsistencies between the model predictions and observations might result from distortions of plasma measurements by extreme heliospheric conditions consisting in very fast solar wind streams (~1000 km/s) and intense fluxes of solar energetic particles. We speculated that anomalous dynamics of the magnetosphere could be well described by the models if the He abundance in the solar wind was assumed to be >20%, which is well appropriate for erupting filaments and which is in agreement with the upper 27% threshold for the He/H ratio obtained from Cluster measurements.

## 1. Introduction

During recent years, great attention was paid to events of extreme magnetospheric disturbances [*Tsurutani et al.*, 2003; *Gopalswamy et al.*, 2005; *Baker et al.*, 2013]. Those unusual events could be characterized not only by extremely strong *Dst* variations but also by extremely small size of the magnetosphere during strong or even moderate magnetic storms [*Vaisberg and Zastenker*, 1976; *Lu et al.*, 1998; *Dmitriev et al.*, 2005a].

The shrinking of the dayside magnetosphere can be caused either by erosion under southward orientation of the interplanetary magnetic field (IMF) or by an enhancement of the solar wind dynamic pressure [*Chapman and Ferraro*, 1931; *Spreiter et al.*, 1966; *Fairfield*, 1971; *Gosling et al.*, 1982]. The effect of southward IMF results in saturation that limits the magnetospheric shrinking [e.g., *Suvorova et al.*, 2005; *Dmitriev and Suvorova*, 2012]. The solar wind dynamic pressure (*Pd*) can achieve very high magnitudes of ~100 nPa that cause very strong compression of the whole magnetosphere such that geosynchronous satellites are located temporarily in the magnetosheath or even in the interplanetary medium.

Table 1 presents a list of such extreme events when the bow shock and magnetopause were situated inside geosynchronous orbit, i.e. at geocentric distances below 6.6 Earth's radii (Re). Most of the events were accompanied by northward or alternating IMF. Hence, the extremely small size of the magnetosphere is mainly caused by abnormally high *Pd* of several tens on nPa. Very high pressures are produced by fast and dense solar wind plasma streams, which are characterized by velocities *V* > 700 km/s and densities *D* of several tens of particles per cc.

Such extreme conditions in the solar wind are developed either in strongly compressed sheath regions downstream of fast interplanetary shocks preceding interplanetary coronal mass ejecta (ICME) or inside so-called erupting filaments, which follow ICME and carry out chromospheric material ejected during solar flares [e.g., *Schwenn*, 1983; *Crooker et al.*, 2000; *Foullon et al.*, 2007; *Chen*, 2011]. The erupting filaments are characterized by significant helium abundance, which substantially contributes to *Pd* [*Gosling et al.*, 1980; *Borrini et al.*, 1982].

In Table 1, one can see that two events of very high *Pd* occurred during storm onset. Apparently, they were related to strong compression in the interplanetary sheath region. Other three events of extreme *Pd* were observed on the recovery phase and they might be related to erupting filaments. It was well established that the great pressure enhancement of ~90 nPa at 19 UT on 21 January 2005 was produced by an erupting filament [*Foullon et al.*, 2007]. *Burlaga et al.* [1998] also reported a very high-density ($D$ > 185 $cm^{-3}$) region of prominence material from erupting filament with great He abundance at the rear of the magnetic cloud during the January 11, 1997 magnetic storm. However in the latter event, the total solar wind dynamic pressure did not exceed 70 nPa because of a relatively low solar wind speed, $V$ ~ 400 km/s.

In contrast to other events, the extreme *Pd* enhancement on 21 January 2005 occurred during the main phase of the magnetic storm. The strong compression was accompanied by unusual dynamics of the magnetosphere. The Double Star TC-1 satellite crossed the bow shock and entered the upstream solar wind in the dusk region at a geocentric distance of 8.5 Re from 1853 to 1907 UT [*Dandouras et al.*, 2009]. The very close approach of the flank bow shock to the Earth corresponds to a very small distance (much less than 6.6 Re) to the subsolar magnetopause. *Du et al.* [2008] reported that the storm on 21-22 January 2005 was highly anomalous because the storm main phase developed during northward IMF.

On the other hand, *Kuznetsova and Laptukhov* [2011] and *Troshichev et al.* [2011a,b] regarded the storm on 21-22 January 2005 as a usual phenomenon since it occurred under the influence of a large interplanetary electric field *E*m. The unusual *Dst* dynamics was explained by a great enhancement of the geoeffective *E*m with the initial input from the southward IMF *B*z and the succeeding input from the azimuthal IMF *B*y component against the background of the very high solar wind speed (*V*sw > 800 km/s). In addition, *McKenna-Lawlor et al.* [2010] studied the ring current dynamics and demonstrated a good correspondence between magnetic field prediction by the *Tsyganenko and Sitnov* [2005] model and observations of energetic neutral atoms in the beginning of the storm from 1700 to 1900 UT. During that time, the moderate but extended response of the magnetosphere to the strong disturbance was explained by a long-duration evolution in the orientation of *Bz* under conditions of enhanced plasma sheet density.

In the present study, we focus mainly on the extremely strong enhancement of the solar wind dynamic pressure from 19 to 22 UT on 21 January 2005. We show an anomalous response of the magnetosphere to the extremely high pressure such that the existing models fail to predict the magnetospheric dynamics even under northward IMF. Heliospheric and geomagnetic conditions are presented in Section 2. The size of the magnetosphere is investigated in Section 3. Dynamics of radiation belt is studied in Section 4. The results are discussed in Section 5. Section 6 is Conclusions.

## 2. Heliospheric and geomagnetic conditions

The magnetic storm on 21 January 2005 was caused by an ICME generated by the X7.1/3B solar flare in the north-western quadrant of the solar disk (N14W61) that occurred at ~0640 UT on 20 January 2005 [*Foullon et al.*, 2007]. The flare produced one of the most intense fluxes of relativistic solar energetic particles (SEP) [*Belov et al.*, 2005; *Kuznetsov et al.*, 2005]. Very intensive fluxes of high-energy SEPs resulted in radiation effects in space instruments that lead to distortion of the space data [e.g., *Dmitriev et al.*, 2005b].

Heliospheric and geomagnetic conditions during the storm on 21 January 2005 are shown in Figure 1. The storm started from a sudden commencement observed at 1710 UT when a strong interplanetary shock (IS) pushed the magnetosphere. At the shock, the solar wind velocity enhanced up to ~900 km/s. The shock accelerated protons with energies up to 30 MeV as measured by GOES-10. The peak flux of >30 MeV protons was ~25 $(cm^2\ s\ sr)^{-1}$. Such conditions were close to the threshold of 50 $(cm^2\ s\ sr)^{-1}$ and $V$ ~ 1000 km/s, which was reported for the plasma data distortion at SWEPAM instrument of the ACE upstream monitor [e.g., *Dmitriev et al.*, 2005b]. Hence, we have to consider plasma data from the ACE and other upstream monitors very carefully.

High-resolution (< 1 min) solar wind plasma data were acquired from the ACE/SWEPAM instrument. We also use summary plasma parameters such as density, velocity and temperature provided by the Cluster Hot Ion Analyzer (HIA) instruments from all probes, and densities of low-energy He and protons measured by Composition and Distribution Function analyzer (CODIF) instrument onboard Cluster C4(Tango) [*Rème et al.*, 2001]. IMF data were obtained from the ACE/MAG instrument and the Cluster C3(Samba) flux gate magnetometer (FGM) [*Balogh et al.*, 2001]. Note that magnetic measurements of other Cluster probes were very similar to those provided by Samba. During the storm, the Cluster satellites were located in the interplanetary medium at $X_{GSM}$ ~ 15 Re, $Y_{GSM}$ ~ 12 Re, $Z_{GSM}$ ~ -3 Re, i.e. in the postnoon sector. The time profiles of the ACE and Cluster data are shifted using the time lags for solar wind propagation to the Earth (~ 1 min and around 30 min, respectively).

As one can see in Figure 1, the plasma and magnetic data from the ACE and Cluster satellites are in very good agreement, excepting profiles of *D*, He/H and IMF *B*x during the interval from ~1900 to 2130 UT. It seems that the relatively low He contribution detected by ACE could result from malfunction of the Composition Aperture telescope of the SWEPAM instrument in the very fast solar wind stream (*V* ~ 1000 km/s) and under enhanced fluxes of high-energy SEP as it happened during 29–31 October event [*Dmitriev et al.*, 2005a]. A strong difference in *B*x is revealed during time interval from ~1945 to 2100 UT when ACE observed large negative *B*x while Cluster observed large positive one. *Foullon et al.*, [2007] reported that the solar wind structure with negative *B*x was also observed by the Wind and Geotail satellites located as ACE in the dawn hemisphere. The authors explained the strong difference in *B*x profiles by a tilted and curved current sheet whose center of curvature was in the north-dawn sector while Cluster was located in the dusk sector. In other words, Cluster observed only a part of the solar wind affecting the magnetosphere in the postnoon region. The prenoon and dawnside magnetosphere was affected by a different solar wind structure.

Cluster and ACE observed different magnetic fields and solar wind density *D*. Strong electric currents should exist in the space between them. Those electric currents and dense plasma stroked on the magnetosphere around this time and partially penetrated inside. The specific and unusual case was that the solar wind and IMF parameters were highly inhomogeneous on the scale size of the magnetosphere and distorted its structure.

Therefore, the total solar wind dynamic pressure *Pd* can be calculated separately for the dawn (ACE) and dusk (Cluster) sectors using the following expression:

$$Pd = 1.67 \cdot 10^{-6} DV^2 (1 + 4\frac{He}{H}), \qquad (1)$$

where *D* and *V* were measured by ACE or Cluster, and helium contribution He/H was acquired from the Cluster C4 data. As one can see in Figure 1, the resultant *Pd* are quite different within the interval from ~1900 to ~2200 UT. The solar wind dynamic pressure is further used for correction of the *Dst* index in order to eliminate the effect of Chapman-Ferraro current at the magnetopause and reveal the contribution of inner magnetospheric currents. We apply an expression derived by *O'Brien and McPherron* [2002]:

$$Dst^* = Dst - 8.6\sqrt{S(\psi)}(\sqrt{Pd} - 1.5), \qquad (2)$$

where $S(\psi) = \frac{1.15}{(1+3\cos^2\psi)^{2/3}}$

Here *Dst* and *Pd* are expressed in nT and nPa, respectively, and $\psi$ is subsolar magnetic colatitude.

In the beginning of the storm, from 1712 to 1846 UT, the dynamics of *Dst* and *Dst** indices can be well described as a function of *B*z (actually *E*y=*V*·*B*z) and *Pd*. From 1710 to 1722 UT, *Dst* increased abruptly from –20 to ~ 60 nT due to an enhancement of *Pd* from a few to ~ 20 nPa. Prominent decreases of *Dst* and *Dst** correspond to intensification of the ring current during intervals of southward IMF from ~1720 to 1750 UT and from 1820 to 1840 UT. *McKenna-Lawlor et al.* [2010] reported that the ring current was well developed by 1900 UT. An increase of *Dst* and *Dst** from 1750 to 1820 UT was caused by recovery of the ring current under northward IMF.

We also use empirical models in order to predict the storm-time *Dst* variation. Figure 2 shows a comparison of the observed hourly averaged *Dst* variation with predictions by Wang model [*Wang et al.*, 2003], MO model [*McPherron and O'Brien*, 2001], OM model [*O'Brien and McPherron*, 2000],

FL Model [*Fenrich and Luhmann*, 1998], and Burton model [*Burton et al.*, 1975]. Note that in the *Dst* (ring current) prediction model, the injection only occurs when IMF is southward, and the decay rate may be dependent on *V*, *B*y, *B*z and *Pd* for some models. All models are optimized based on a number of historical data. As one can see in Figure 2, all the models predict *Dst* quite well in the beginning of the storm from 1700 to 1900 UT. However after 1900 UT, all the models fail and predict a recovery phase while the *Dst* decreases sharply on several tens of nT. The decrease could not be predicted by any model because IMF was mainly northward during that time.

In Figure 1 one can see that most prominent difference between *Dst* and *Dst** is revealed from ~1900 to 2100 UT. During that time, *Dst* was decreasing by ~85 nT from ~45 nT to ~ -40 nT, while *Dst** corrected by the ACE pressure was almost constant and varying around –70 nT. Hence, the dynamics of *Dst* can be well attributed to a decrease of *Pd* from ~120 to 20 nPa. However, behavior of *Dst** is anomalous because under positive *Bz*, the ring current should decay and, thus, *Dst** should increase. It looks like the ring current did not decay from ~1900 to 2100 UT.

From 2055 to 2115 UT, *Dst** has decreased from ~ -70 to ~ –130 nT. The strong decrease of *Dst** is hard to explain by short intervals with negative *B*z of small magnitude as well as by variations in *Pd*. From 2115 to 2400 UT, IMF remained northward and *Pd* was varying about 30 nPa. During this time, *Dst** started to increase that indicates to decay of the ring current. However, this decay was abnormally slow.

## 3. Geosynchronous crossings of the magnetopause and bow shock

We determine the size of the magnetosphere using observations and modeling of the magnetopause and bow shock by geosynchronous satellites. The magnetopause is modeled by an empirical model of Kuznetsov and Suvorova [*Kuznetsov and Suvorova*, 1998; *Suvorova et al.*, 1999] (hereafter KS98 model), which has demonstrated very good capabilities for prediction of the dayside magnetopause in very wide dynamic range and enables predicting a storm-time dawn-dusk asymmetry [*Dmitriev et al.*, 2005a; 2011]. Note that KS98 model demonstrates best capabilities in prediction of the strongly compressed magnetopause under northward IMF [*Suvorova et al.*, 2005]. We also use an empirical model by *Dmitriev et al.* [2011] predicting the solar wind pressure *P*gmc required for geosynchronous magnetopause crossing at a given location. Namely, if *P*gmc is lower (higher) than *Pd* then a geosynchronous satellite is expected to be located in the magnetosheath (magnetosphere). This model is based on advanced set of geosynchronous magnetopause crossings observed in an extremely wide range of IMF *Bz* from –30 to 30 nT.

The bow shock is modeled by *Verigin et al.* [2001] model (hereafter BSV model) and by *Chao et al.* [2002] model (hereafter BSC model). Note that the BSV model depends on the size and shape of the dayside magnetopause, which is calculated by the KS98 model [e.g., *Dmitriev et al.*, 2003]. The BSC model does not depend on modeling of the magnetopause. The BSV and BSC models demonstrated quite high capabilities for prediction of the bow shock in the previous statistical studies [*Dmitriev et al.*, 2003].

We also use results of global MHD modeling of the magnetosphere performed by SWMF/BATS-R-US code with Fok ring current (version v20110131) provided by the Community Coordinated Modeling Center (Alexei_Dmitriev_072512_1). The model is driven by upstream solar wind and IMF data acquired from the ACE satellite within the time interval from 1630 to 2400 UT on 21 January 2005. The code allows tracing of geosynchronous and other satellites to obtain model values of magnetic field and plasma parameters along the orbit.

Figure 3 shows the location of GOES-10, GOES-12, LANL-1990, LANL-1994, LANL-1997, Cluster and Double Star TC-1 satellites at ~1850 UT on 21 January 2005. The profiles of magnetopause and bow shock are calculated, respectively, by the KS98 and BSV models for extreme solar wind conditions: Alfven Mach number *M*a = 8, sonic Mach number *M*s = 12, *B*z = -20 nT, *Pd* = 90 nPa. As one can see, the subsolar bow shock and practically the whole dayside magnetopause are located inside geosynchronous orbit such that all the geosynchronous satellite should be located either in the magnetosheath or even in the upstream solar wind.

Figure 4 shows GOES-10 and GOES-12 observations of the magnetic field from 17 to 24 UT on 21 January 2005. The magnetopause crossed GOES-12 at local noon right in the beginning of the storm at 1712 UT. Until 1840 UT, GOES-12 was located in the magnetosheath where *B*x, *B*y and *B*z components of the magnetic field were strongly magnified and correlated well with the corresponding IMF components observed by Cluster. At the same time, GOES-10 was located in the dawn – prenoon sector and encountered with the magnetosheath from 1736 UT to 1750 UT and from 1821 to 1846 UT.

At 1846 UT, both GOES-10, located in the prenoon sector, and GOES-12, located in the postnoon sector, crossed the bow shock and came into the interplanetary medium where they observed practically the same magnetic field as Cluster. The satellites situated upstream of the bow shock during ~2 hours and returned to the magnetosheath at 2035 and 2010 UT, respectively. That long duration of the interplanetary interval is really outstanding for the geosynchronous satellites.

During the interplanetary interval, the GOES satellites observed positive IMF *B*x, which was consistent with the Cluster observations. Note that at 1945 UT, ACE observed a reversal of the IMF *B*x component (see Figure 1). Hence, it is reasonable to suggest that Cluster observed the solar wind and IMF conditions, which did affect most part of the dayside magnetosphere from the prenoon (GOES-10) to dusk (GOES-12) sector.

In Figure 4 one can see that from 1712 to 1846 UT, the magnetopause crossings and magnetosheath intervals are well predicted by KS98 and MHD models both in prenoon and postnoon sectors. The dynamics of modeled pressure *P*gmc is also in good agreement with the observations: time intervals of *P*gmc < *Pd* correspond well to the magnetosheath intervals observed by the GOES satellites.

However, the interplanetary interval from 1846 to 2035 UT cannot be completely predicted by the models. The BSV model predicted only a brief solar wind encounter from 1846 to 1855 UT when the IMF turned southward. The BSC model, based on either ACE or Cluster dynamic pressure, cannot predict any bow shock crossings. The MHD model predicts strong variations of high-amplitude magnetic field that rather typical to the magnetosheath than to the interplanetary magnetic field. Hence, the models fail to predict the bow shock location for the present event.

Additional inconsistencies can be found during the GOES-12 magnetosphere encounter at 2130 UT when the solar wind dynamic pressure was decreasing gradually. The magnetopause crossing was observed under decreasing *Pd*, which was already much lower than *P*gmc for ~20 min. The KS98 model also predicted the magnetopause crossing much earlier (at ~2105 UT) than actual one. However, the magnetospheric encounter by GOES-10 at 2340 UT was predicted by KS98 quite precisely.

In Figure 5, we show magnetosheath interval observed by LANL-1997 from 1912 to 2400 UT. LANL satellites do not detect magnetic field but they measure plasma characteristics. For this case, the magnetopause crossings are identified by using so-called a ratio of ion density to temperature (RI) and of electron density to temperature (RE) (see details in *Suvorova et al.,* [2005]). In the hot magnetospheric cavity, the ratios RI and RE are small (< 1) while they are high (~100) in the dense and hot magnetosheath. Note that the actual threshold can become lower due to a radiation effect of SEP [*Dmitriev et al.*, 2005a]. For the present case we use the threshold of RI ~ 10.

At 1912 UT, LANL-1997 crossed the magnetopause and encountered with the magnetosheath at very early local morning (~0530 MLT). The magnetopause crossing by LANL-1997 was in good agreement with the value of *P*gmc, which was smaller than *Pd* measured by Cluster. However, the KS98 model could not completely predict the magnetosheath interval. The model overestimated the magnetopause distance from 1912 to ~2100 UT. Hence, we can conclude that the KS98 model fails to predict the magnetopause crossings during time interval from 1912 to 2130 UT both in the dawn and dusk sectors. It seems that higher *Pd* is required for the KS98 model in order to give a correct prediction for GOES-12 and LANL-1997.

From the observations, we can determine that in the noon region, the magnetopause was located inside geosynchronous orbit from 1712 on 2400 UT. The minimal distance to the magnetopause of ~3 Re was predicted by KS98 model at ~1850 UT. From 1846 to 2035 UT, geosynchronous orbit in the noon region was located upstream of the bow shock and practically whole dayside magnetopause came inside geosynchronous orbit. We can approximately estimate the magnetopause distance during the interplanetary interval taking into account an average ratio of 1.3 for the distances to the subsolar bow shock and magnetopause [*Spreiter et al.*, 1966]. For the bow shock distance of 6.6 Re we obtain the magnetopause distance of ~5 Re. Hence during almost 2 hours from 1846 to 2035 UT 21 January 2005, the magnetosphere was extremely compressed such that the distance to the subsolar magnetopause was less than 5 Re.

**4. Dynamics of the ring current and radiation belt**

Extremely strong long-lasting compression of the magnetosphere should affect the radiation belt and dynamics of the ring current. The fast and dramatic magnetosphere shrinking from 1846 to 1855 followed by an ~2-hour decrease of the compression should violate the third adiabatic invariant of protons with energies from tens of keV to a few MeV in the outer magnetosphere at drift shells $L > 4$. Therefore from 1846 to 2035 UT 21 January 2005, the radiation belt and ring current should be significantly modified and restricted by the upper boundary located at $L \sim 5$. Here we use low-orbit high-inclination satellites CORONAS-F and POES for studying the radiation belt and ring current dynamics.

Figure 6 shows temporal variations of pitch-angle anisotropy for the protons with energies of tens of keV observed by POES-17 near the noon-midnight meridian on 21 January 2005 at $L \sim 5$ corresponding to the outer magnetosphere. The anisotropy is calculated as a ratio between trapped proton fluxes with pitch angles $\alpha \sim 90°$, i.e. perpendicular to the magnetic field line, to precipitating ones with $\alpha \sim 0°$. Before the magnetic storm, the satellite observed mostly trapped energetic protons gyrating near their mirror points such that the ratio was varying around 100. During magnetospheric compression at ~18 UT and especially from ~1900 to ~2200 UT, the anisotropy was mainly ~1 and even less than 1 that corresponded to a diminishing the trapped proton population in the outer magnetosphere.

Dynamics of proton fluxes observed by POES-17 satellite near the noon-midnight meridian on 21–22 January 2005 is shown in Figure 7. Before the storm at 1700 UT, integral fluxes of low-energy (> 30 keV) protons had a maximum of up to ~$10^9$ $(cm^2\ s\ sr)^{-1}$ at $L$ = 4. During the storm development, the fluxes were substantially increasing mainly in the inner magnetosphere at $L$ < 4 such that at 2100 UT, the fluxes of >30 keV, >80 keV and > 200 keV protons enhanced by almost two orders of magnitude and peaked at $L$ = 2 and 3. In contrast, the proton fluxes have diminished in the outer magnetosphere at $L$ = 4 and 5. Such dynamics corresponds to fast transport of the ring current particles into the inner regions and losses of radiation belt particles at $L$ > 4 (magnetopause shadowing) due to a strong and long-lasting compression of the magnetosphere.

CORONAS-F satellite observed a similar dynamics of energetic protons (1–5 MeV) as shown in Figure 8. From ~18 to ~22 UT, the fluxes in the inner magnetosphere increased up to 3 orders of magnitudes. Most significant proton enhancement can be revealed in the range of $L$-shells from 2 to 4. It is important to note that the proton fluxes at $L$ = 2–3 have diminished very fast after 23 UT that is caused by very intense particle losses in the inner magnetosphere. Note that at $L$ = 3–5, the particle fluxes remained high and were decreasing gradually during the rest of the storm.

From observation of the low-energy protons we found that the extreme compression of the magnetosphere from ~1850 to ~2100 UT on 21 January 2005 was accompanied by anomalous transport of the particles from the outer to the inner regions. The outer magnetosphere at $L$ > 5 was pushed inward during the extreme compression. The particles from the radiation belt and ring current were accumulated in the inner magnetosphere at $L$ < 4. The dynamics of the proton fluxes in the inner magnetosphere did not reveal substantial losses until the end of compression.

## 5. Discussion

From analysis of the geomagnetic storm on 21 January 2005, we have found that the storm can be divided in two parts accompanied by essentially different solar wind dynamic pressures. The beginning of the storm lasted from 1712 to 1846 UT under *Pd* < 20 nPa. During this phase, the dynamics of the magnetospheric boundaries, magnetopause and bow shock, as well as the ring current are well predicted by empirical and first-principle models. The situation changed dramatically after 1846 UT when an extremely high solar wind pressure and strong southward IMF (at 1846–1855 UT) smashed out the outer magnetosphere such that a part of geosynchronous orbit at 10 to 14 MLT occurred inside the upstream solar wind for almost 2 hours.

We can make indirect estimation of the subsolar distances for the extremely compressed magnetopause and bow shock on the base of the fact that from 1853 to 1907 UT, Double Star TC-1 entered into the upstream solar wind [*Dandouras et al.*, 2009]. We use various models [see *Dmitriev et al.*, 2003 for details] in order to calculate the bow shock subsolar distances *R*s when Double Star TC-1 crosses the bow shock at GSM location (*X* = 1.3, *Y* = 7.4, *Z* = 4.0 Re) under strong southward IMF (*B*z = -23 nT). We also use a model shape of the bow shock proposed by *Cairns et al.* [1995]. Table 2 shows the resultant *R*s and *Pd* required for the Double Star TC-1 crossing of the bow shock. Only models by *Russell and Petrinec* [1996], BSV and BSC enable to predict the crossing for the given solar wind conditions. Other models overestimate the bow shock distance substantially. From Table 2 we find that from 1853 to 1907 UT, the subsolar bow shock was located below 5.2 Re and, thus, the magnetopause nose distance was smaller than 4 Re. Note that actual distances to the bow shock and magnetopause could be much smaller.

During the period of extreme magnetospheric compression, the behavior of the magnetosphere became very unusual: all the models failed to predict the magnetospheric dynamics. Namely, no model could predict the extremely small size of the magnetosphere: bow shock location at 6.6 Re for ~2 hours and magnetosheath encounter at very early local time of ~0530 MLT. The empirical models could not predict the anomalous increase of negative *Dst* variation, or storm main phase, observed under northward IMF that meant an unusual intensification of the “non-decaying” ring current. It seems that the models may be not workable for extreme condition such as extremely compressed magnetosphere or steady northward IMF. In addition, the *B*y component of IMF is large for this event. There may exist partial component magnetic reconnection at the subsolar point when there is *B*y component. This may also contribute to injection of ring current particles as proposed by *Kuznetsova and Laptukhov* [2011] and *Troshichev et al.* [2011a,b].

*Du et al.* [2008] proposed two possible mechanisms to explain the anomalous behavior of *Dst*. The first one consists in a lengthy storage of solar wind energy in the magnetotail and delayed release into the ring current. However, we do not find any particle injections in the outer magnetosphere during time interval from 1999 to 2035 UT. Instead, in the night and evening sectors, we observe weaker fluxes at $L$ = 4–5 than those at $L$ = 3–4. Decreases of *Dst* after 2035 UT might be caused by intensification of the substorm activity observed under strong compression in the sub-auroral zone [*Lazutin and Kuznetsov*, 2008; *Lazutin et al.*, 2010]. The substorm activity was caused by enhancements of the solar wind dynamic pressure and southward IMF turnings observed by the satellites ACE, Cluster (See Figure 1) and GOES-10 (see Figure 4).

The second mechanism proposed by *Du et al.* [2008] that during the storm, the plasma sheet may be close to the Earth, resulting in a large contribution of the tail current to the *Dst* index. However, the inner part of tail current, being strong and close to the Earth in the beginning of compression, should move out and become weaker within ~10 minutes after a decrease of the solar wind dynamic pressure and northward IMF turning [*Borovsky et al.*, 1998; *Tsyganenko*, 2000]. The magnetic effect of tail current was found to be dominant in the *Dst* variation during moderate magnetic storms with $Dst_{\min}$ > -100 nT [*Ganushkina et al.*, 2010]. As shown by *Tsyganenko* [2000], the best driving parameters for the tail current are lg(*Pd*) and a complex function of the solar wind velocity $V$, IMF transversal component $B_\perp=(By^2 + Bz^2)^{1/2}$ and clock angle θ: $\varepsilon = V\cdot\sin^3(\theta/2) \times (B_\perp/B_c)^2/(1+B_\perp/B_c)$, where $B_c$ = 40 nT. In Figure 9 one can see that from 1845 to ~1900 UT, both ε and *Pd* increase dramatically and, thus, the tail current contribution to negative *Dst* was significant at that time. However, after 1900 UT, both ε and *Pd* decrease rapidly that indicates to diminishing tail current. Hence, the tail current cannot explain the “non-recovering” *Dst*.

Another possible source of the ring current might be solar energetic particles [*Hudson et al.*, 1997; 2004; *Richard et al.*, 2009]. It has been shown that SEP penetration is effective during strong compression of the magnetosphere by interplanetary shocks. However, the SEP flux during the shock passage at ~1845 UT was not very strong (~$10^3$ $(cm^2\ s\ sr)^{-1}$ for >1 MeV protons as shown in Figure 1) such that the SEP protons could contribute only a little portion of the ring current. Further after the compression, trapped and quasi-trapped particles are lost by motion through the magnetopause and by precipitation. This should result in a gradual decrease of the particle fluxes and, thus, a decrease of their contribution into the ring current. Hence, we can neglect the effect of SEP penetration.

The mechanisms proposed cannot also explain the observations of both extreme and long-lasting magnetopause compression. The negative magnetic effect to the subsolar geomagnetic field (if any) should diminish with decreasing *Pd* and the magnetosphere should expand as predicted by the models during time interval from ~19 to 21 UT on 21 January. However, we did not find this expansion in both the bow shock location and radiation belt profile. Instead, the standoff magnetopause was below 5 Re during that time.

Here we have to remind that the solar wind plasma of very high density was originated from an erupting filament [*Foullon et al.*, 2007], which usually contains a significant portion of He. *Sharma et al.* [2013] reported very large He to proton ratio of >20% in the filament plasma observed by ACE/SWICS on 7–8 January 2005. It is important to note that those days were not accompanied either by enhanced SEP fluxes or by very fast solar wind and, hence, the Composition Aperture telescope onboard ACE was operating safely. In contrast during the 21 January storm, the SEP fluxes were intense and the solar wind speed was high (see Figure 1) such that both ACE and Cluster detectors suffered from the radiation impact [*Foullon et al.*, 2007]. The two satellites measured very similar proton density but very different He to proton ratio. Hence, it is reasonable to assume that the experimental data on the He/H ratio are not reliable and He contribution can be underestimated.

Figure 10 demonstrates Cluster plasma data from the more recent calibrations of the CIS team (communicated to us by the anonymous Reviewer of this paper). It is well known that CODIF is a time-of-flight ion mass spectrometer, designed mainly for magnetospheric ions, and it can thus be saturated under intense solar wind fluxes as those encountered here (*Rème et al., 2001*; and *CIS User Guide*, available at the CAA: http://caa.estec.esa.int/caa/ug_cr_icd.xml). HIA, in the low-sensitivity side operation (as was the case here), can instead handle very intense fluxes without this saturation problem. This is evident from Figure 10 where after 1844 UT, HIA measured a jump of the solar wind density up to ~57 $cm^{-3}$, whereas CODIF at this time showed no increase of the proton density, and showed even a small "decrease", typical for saturation conditions. The total ion density is thus supplied by HIA. CODIF, however, can still give a rough measure of the $He^{++}$ contribution. As shown in Figure 1, the "measured", under saturation conditions, proton density was ~8.5 $cm^{-3}$, whereas at the same time the "measured" $He^{++}$ density was ~2.3 $cm^{-3}$. This gives a $He^{++}$ to proton ratio of ~27%. Note that this is an upper limit, because the proton channel suffers from a stronger saturation than the $He^{++}$ channel (due to the much higher proton fluxes, as shown in Figure 10, indicating stronger saturation signatures). The actual $He^{++}$ to proton ratio is thus clearly less than 27% but above the ~8% shown as the "observed" CODIF He/H ratio in Figure 1.

Therefore the discrepancies between the observations and model predictions can be originated from "insufficiently strong" solar wind dynamic pressure because of underestimation of the He contribution. Using empirical models of the bow shock and magnetopause, we can estimate the He contribution and *Pd* required for observed magnetopause and bow shock crossings. In Figure 9, we show predictions of the empirical models for a "synthetic" *Pd* derived from the Cluster HIA measurements but with 4-time magnification of the He contribution acquired from original Tango/CODIF plasma measurements (see Figure 1).

The "synthetic" *Pd* is very close to the observed one in the beginning of the storm because of very low original He content at that time. During the extreme compression, the He/H ratio increased up to ~30% and, thus, the "synthetic" solar wind dynamic pressure enhanced up to 200 nPa. As one can see in Figure 9, the magnitude and dynamics of "synthetic" *Pd* allows successful predicting the interplanetary interval from 1846 to 2035 UT. Moreover, the profile of pressure corrected *Dst** becomes not so much anomalous. Namely, a decrease from 1846 to ~1920 UT can be attributed to intensification of the ring current due to the southward IMF turnings observed at that time by both ACE and Cluster satellites.

Helium contribution of ~30% shown in Figure 9 is slightly higher than 27% upper threshold obtained from Cluster measurements. The 10% difference leads to ~5% decrease in the solar wind dynamic pressure from 195 to 184 nT. This decrease is small and does not affect much the results obtained above. Actual value of the He/H ratio could be even smaller than 27% leading to the decrease in *Pd* of ~ 20% and even 30%. However in the KS98 model, the size of magnetopause depends on the dynamic pressure as $Rs \sim (Pd)^{-1/5.2}$ and the pressure correction of *Dst* depends on $Pd^{-1/2}$. Hence, the decrease of *Pd* gives the result lying within the model errors, especially in the range of extrapolation. Much more important problem, especially for the *Dst* correction, is temporal dynamics of the He/H ratio, which is hard to derive from the experimental data.

Based on the "synthetic" data we obtain that from 1920 to 2035 UT, *Dst** was varying around -100 nT and did not practically decrease despite of northward IMF. This effect can be related to the dynamics of ring current during the strong magnetospheric compression. The trapped particles were moving to lower *L*-shells and accelerated adiabatically in a betatron mechanism, which was keeping the first two adiabatic invariants. This process enforced the ring current. The abrupt increase and long lasting decrease of magnetospheric compression of duration comparable with the drift periods of particles in the outer zone resulted in violation of the third adiabatic invariant. Hence after the extreme compression, the particles gained energy and remained at lower L-shells. This effect can be revealed in Figure 7 as a strong increase of the low-energy proton fluxes observed by POES in the inner magnetosphere at ~21 UT, i.e. in the end of the extreme compression. In contrast, there is a deficiency of protons in the outer magnetosphere.

Qualitative estimations of the effects of adiabatic transport and intensification of the ring current are conducted in Appendix A. We found that the low energy protons were accumulated and kept high fluxes in the inner magnetosphere at $L < 4$ such that the total number of particles in the ring current did not practically change. However, the inner magnetosphere is characterized by very intense losses of the low-energy protons in charge-exchange interactions with neutral atoms of the exosphere [see *Cornwall and Schulz*, 1979; *Kistler et al.*, 1989]. Such losses should result in a fast decay of the ring current and formation of recovery phase with positive variation in *Dst**, which was not observed.

Exosphere's density at $L > 3.5$ varies within 20% and increases during magnetic storms [e.g., *Østgaard et al.*, 2003; *Bailey and Gruntman*, 2013], that promotes a decrease of lifetime of ions in the ring current. In addition, statistical studies of magnetic storms found that the decay time decreased with increasing solar wind dynamic pressure [*Wang et al.*, 2003]. It was also shown that the decay time during recovery phase depends on the storm magnitude: for moderate storms with $Dst_{\min} > -125$ nT, the decay time increases with the storm magnitude [*Pudovkin et al.*, 1985], while for strong magnetic storms an opposite effect was revealed [*Feldstein et al.*, 1984].

The charge-exchange decay of the ring current and decay time dependence on the ring current location are controlled by two concurrent effects [*Kovtyukh*, 2001]: (1) the closer location of the ring current the higher exosphere's density that decrease the decay time; (2) with moving toward the Earth, ring current particles suffer betatron accelerated and if the cross-section for charge-exchange decreases with increasing energy, then the lifetime of ring current ions should increase. For the protons with energies $E > 30$ keV, the cross-section for charge-exchange decreases fast [*Claflin*, 1970; *Cornwall and Schulz*, 1979; *Kistler et al.*, 1989] and, hence, the latter effect dominates: the decay of the ring current decreases with decreasing distance to the Earth. For oxygen ions ($O^+$), the charge-exchange cross-section has a dependence on the energy much weaker than that for the protons [*Cornwall and Schulz*, 1979; *Kistler et al.*, 1989] that results in dominance of the first effect: the decay of ring current increases with decreasing the distance to the Earth.

During strong magnetic storms enriched by $O^+$ the decay time is short in the beginning of recovery phase [see *Kovtyukh*, 2001 and references therein]. During weak and moderate storms, such as the event considered, the ring current is manly contributed by protons. Hence, shifting the ring current toward the Earth (as considered in details in Appendix A) results in significant increase of its lifetime. Therefore, we can propose an effect of weak particle losses in the inner magnetosphere.

We can estimate the change of the lifetime for the protons in the maximum of ring current during its energization and earthward shifting. The lifetime can be calculates as $\tau = (\sigma \nu n_H)^{-1}$, where σ is the cross-section for charge exchange, *v* is velocity of protons and $n_H$ is density of the exosphere. From Appendix A we find that the maximum of ring current moves from *L* ~ 4.4 to *L* ~ 3.1 and, thus, the exospheric density increases by ~3 times from 200 to 600 $cm^{-3}$ [*Østgaard et al.*, 2003]. At the same time, the protons are accelerated when move to the region with higher magnetic field. The acceleration can be estimated as a ratio of magnetic field strength at *L* ~ 3.1 and *L* ~ 4.4 near equatorial plane: $(4.4/3.1)^3$ ~ 3. Hence, the proton velocity increases as $\sqrt{3} \sim 1.7$. In the energy range above 30 keV, the cross-section of charge-exchange for protons decreases with particle energy *E* approximately as $E^{-4}$ [*Claflin* 1970]. If the proton energy increases by 3 times then the cross-section decreases by ~80 times. Therefore, the lifetime of protons increases by 80/(1.7·3) > 10 times and, thus, accelerated protons of the ring current can survive in the inner magnetosphere for a long time and support the "non-decaying" ring current from 1920 to 2035 UT.

During the extreme compression, the outer magnetosphere (*L* > 5) was affected by a dense and slow plasma fluxes from the magnetosheath. The solar wind flux can be estimated as $j_{sw} = V \times D$ = 940 km/s × 50 $cm^{-3}$ ~ $5 \cdot 10^9$ $(cm^2\ s)^{-1}$. This flux affects a large portion of the outer magnetosphere on the dayside during *t* ~ 2 hours (~ 7000 s). In the magnetosheath, the solar wind ions are decelerated to ~ 1 keV energies, which correspond to the cross-section for protons charge-exchange of σ ~ $2 \cdot 10^{-15}$ $cm^2$ [*Claflin*, 1970]. We can roughly estimate the relative decrease of the exospheric density as $\sigma \cdot j_{sw} \cdot t$ ~ $7 \cdot 10^{-2}$. Hence, almost 10% of the outer exosphere is eroded by the solar wind that results in ~10% increase of the lifetime of the ring current populating the outer magnetosphere after the decrease of solar wind pressure at ~2035 UT. This effect might also contribute to very slow recovery of *Dst** after 2115 UT.

Finally, we have to emphasize that accurate pressure correction of the *Dst* variation is crucially important for estimation of the ring current contribution to the storm-time magnetosphere dynamics [e.g., *Lu et al.*, 1998]. However during most of extreme events, we can not get completely reliable data on the key plasma parameters, especially proton density and He contribution, because of very fast solar wind streams and/or SEP radiation impact to the plasma instruments [*Dmitriev et al.*, 2005b; *Foullon et al.*, 2007; *Russell et al.*, 2013]. Design of new space plasma instruments robust under extreme conditions should become an important issue for the future space missions.

**6. Conclusions**

Analysis of the solar wind conditions and dynamics of the magnetosphere and radiation belt during anomalous magnetic storm on 21 January 2005 has provided us the following findings:

1. The storm was unusual because it was developing under very strong solar wind dynamic pressure and/or large northward IMF such that from 1712 to 2400 UT, the noon region of geosynchronous orbit was continuously located in the magnetosheath and was exposed to the upstream solar wind during ~2 hours.
2. The beginning part of the storm, lasting from 1712 to 1846 UT, was typical and can be successfully predicted by the existing models of the magnetopause and ring current as well as by the global MHD simulations.
3. Anomalous magnetospheric dynamics, under which all the models failed, was revealed after 1846 UT and related to the extremely strong dynamic pressure $Pd > 100$ nPa produced by the dense and fast plasma of erupting filament.
4. During ~2 hours from 1846 to 2035 UT, the outer magnetosphere at $L > 5$ was eliminated. The subsolar bow shock was located inside geosynchronous orbit at distances <6.6 Re that corresponded to the magnetopause standoff distance <5 Re.
5. The ring current dynamics under the extreme compression can be qualitatively described by the earthward transport with adiabatic betatron acceleration accompanied by violation of the third adiabatic invariant that resulted in accumulation of the particles in the inner magnetosphere at $L < 4$.

The lifetime of accelerated protons in the inner magnetosphere is obtained to be >10 times longer than typical one that explains “non-decaying” ring current observed after 19 UT.

6. We speculate that the anomalous dynamics of extremely compressed magnetosphere can be well described by the models if we accept the He abundance of ~30%, which is only slightly higher than the upper 27% threshold for He/H ratio obtained from Cluster measurements that is not unusual for erupting filaments. High helium abundances of >20% allow successful predicting the observations by the empirical models within the model errors.

## Appendix A

### On contribution of the ring current adiabatic compression to the negative Dst* variation

During almost 2 hours from 1845 to 2055 UT, we observed an abrupt and very strong increase preceding a gradual and slow decrease of the solar wind dynamic pressure (see Figure 1). Such kind of pressure variation resonates with the drift periods of ring current (RC) particles (~ 2 hours). This giant pressure pulse resulted in irreversible shift of RC to the Earth. The particles of RC were accelerated in a betatron mechanism with keeping the first two adiabatic invariants of the drift motion that caused an intensification of RC and negative variation in *Dst**. The giant pressure pulse should affect the ring current in the same way as a sudden impulse affects the radiation belt during a storm sudden commencement. From this, we can estimate the magnetic effect produced by the RC intensification.

We will base the calculations on the values measured at $t_1$ ~ 1845 UT on 21 January 2005, i.e. right before the beginning of extreme compression that can be attributed to the end of main phase and beginning of recovery phase of a magnetic storm, and at $t_2$ ~ 21.20 UT 21 January 2005, i.e. immediately after the extreme compression and in the beginning of recovery phase of the following storm. Thereby, these two moments can be attributed to recovery phases of overlapping magnetic storms. This approach greatly simplifies all calculations without specifying the unit system and constant numerical factors, including the coefficients of the integrals. These coefficients are simply reduced in the ratios presented below. Hereafter, unprimed quantities will refer to the time moment $t_1$, and primed – to the time moment $t_2$.

We will compare the deviations of relations calculated for the magnetic effect of RC, which is closed in a trap and gradually compressed, with the following ratios:

$$\frac{D_{st}'}{D_{st}} \approx \frac{80 \pm 10}{40 \pm 5} \approx 2.0 \pm 0.5, \tag{A1}$$

$$\frac{D_{st}^{*\prime}}{D_{st}^{*}} \approx \frac{115 \pm 15}{60 \pm 5} \approx 1.9 \pm 0.4, \tag{A2a}$$

$$\frac{D_{st}^{*\prime}}{D_{st}^{*}} \approx \frac{135 \pm 15}{60 \pm 5} \approx 2.3 \pm 0.4. \tag{A2b}$$

The latter two ratios are written for the *Dst* variation corrected, respectively, on the pressure without He contribution (see Figure 1) and on the “synthetic” pressure with strong He abundance (see Figure 9), which also contributes to the magnetosheath population.

We believe that at moments $t_1$ and $t_2$, the contribution of the tail current was negligibly small (see Figure 9) and RC was quasistationary. Hence, we can apply the Dessler-Parker-Sckopke theorem and get:

$$\frac{D_{st}^{*\prime}}{D_{st}^{*}} = \frac{W'}{W}, \tag{A3}$$

where $W$ – total kinetic energy of all particles in RC. Taking into account all methodic uncertainties, we can consider only ion contribution to *Dst* and neglect a contribution of electrons. From statistical consideration by *Kovtyukh* [2010], we can estimate that on average during the recovery phase, the maximum of RC is located at $L_{\mathrm{m}} \approx 4.4 \pm 0.3$ under $Dst \approx -(40 \pm 5)$ nT, i.e. before the extreme compression, and at $L_{\mathrm{m}} \approx 3.1 \pm 0.2$ under $D_{st} \approx -(80 \pm 10)$ nT, i.e. after the extreme compression.

During the interval between $t_1$ and $t_2$, the noon magnetopause was located very deep inside geosynchronous orbit. However at the moments $t_1$ and $t_2$, the magnetopause was quite close to geosynchronous orbit. Since the outer edge of RC is steep enough, we suppose for definition that at the beginning and end of the interval, the outer boundary was located at $L_{\mathrm{b}} \approx 6.6$. With a more realistic position of the boundary and its offset in $L$ during this period, our simplification has an uncertainty within 10%. That is considerably less than errors related to the uncertainty in localization of the RC maximum. Since the inner edge of RC is always much steeper than the outer one, we can neglect the contribution of particles in the inner edge to the RC total energy.

The radial profile of pressure (energy density) of the hot magnetospheric plasma from the RC maximum to the outer edge of geomagnetic trap can be well approximated by the following expression: $p(L) \approx a\, L^{-2} \exp\left(-L/L_0\right)$, where $L_0 \approx 2$ during recovery phase of magnetic storms [*Kovtyukh*, 2010]. The normalization parameter $a$ varies from storm to storm. The ratio of the parameter values during the two time moments is equal to a ratio of RC pressures at those moment at the same $L$-shell, for instance at $L = 5$. As a rule, in the end of main phase and in the beginning of recovery phase, the ring current is quickly (within ~ 1 hour) symmetrized by MLT. Therefore, we can expect that at the moments $t_1$ and $t_2$, the ring current was almost symmetrical. Hence, in our calculations we suppose a symmetrical RC with isotropic pitch-angle distribution for simplicity.

In the event considered, we can write the following equation for the dipole trap:

$$\frac{W'}{W} = \frac{p'(L=5)}{p(L=5)} \times \frac{\int_{3.1\pm0.2}^{6.6} p'(L) L^2 dL}{\int_{4.4\pm0.3}^{6.6} p(L) L^2 dL} \qquad \text{(A4)}$$

or

$$\frac{W'}{W}=\frac{p'(L=5)}{p(L=5)}\times\frac{\int_{3.1\pm0.2}^{6.6}\exp\left(-L/L_0\right)dL}{\int_{4.4\pm0.3}^{6.6}\exp\left(-L/L_0\right)dL}\approx\frac{p'(L=5)}{p(L=5)}\times(2.5\pm0.8)\,. \tag{A5}$$

In order to adjust (A5) to (A2), we have to suppose

$$\frac{p'(L=5)}{p(L=5)}\approx 0.76-0.92\,. \tag{A6}$$

That is consistent both with the idea of RC compression and earthward displacement during the given time interval and with the RC pressure values at $L$ = 5, which have been obtained for other storms of similar strength [see Table 2 in *Kovtyukh*, 2010].

Further, we calculate a relative change of the total number of RC ions ($N$), which satisfies the inner and outer boundaries of RC accepted here and relative increase of the RC pressure. To do this, we have to select the shape of energy spectra of ions or the shape of the energy dependence for the ion energy density (pressure). In according to experimental data [e.g., *Fritz et al.*, 1974] the latter relationship can be approximated by a Maxwellian distribution, such that the energy density ($p$) and the concentration of hot plasma ($n$) are connected by a well-known simple relation:

$$p(L)\propto E_m(L)\,n(L),$$

where $E_m(L)$ is the location of maximum in the differential Maxwellian distribution. For simplicity, we suppose that the value of $E_m$ changes with $L$ adiabatically, i.e. $E_m\propto L^{-3}$, and $E_m'$ = $E_m$ at $L$ = 5. Then

$$\frac{N'}{N}=\frac{p'(L=5)}{p(L=5)}\times\frac{\int_{3.2\pm0.2}^{6.6}L^3\exp\left(-L/L_0\right)dL}{\int_{4.4\pm0.3}^{6.6}L^3\exp\left(-L/L_0\right)dL} \tag{A7}$$

or, after calculation of the integrals, we get:

$$\frac{N'}{N} = \frac{p'(L=5)}{p(L=5)} \times (1.37 \pm 0.1). \tag{A8}$$

Assuming that the ratio of pressures is the same as that in (A6), we can derive from (A8):

$$\frac{N'}{N} = 1.16 \pm 0.19. \tag{A9}$$

It means that during the extreme compression, a small amount of particles could be injected in the ring current.

In our calculations, the energy spectrum of RC ions could be approximated by more realistic function. However, it greatly complicated the calculations and made a little difference in the results. Under adiabatic compression of particles in quasi-dipole trap, the anysotropy of particle fluxes increases if mechanisms of fast isotropisation are absent. In additioon, azimuthal asymmetry of RC could be changing during the period considered and could be different in the beginning and in the end. The difference could result in some changing in estimations (A6) and (A9) but could not change, apparently, the basic qualitative conclusions.

Thus, our calculations show that under simple assumptions, the change of *Dst* and *Dst**, observed from the moment right before the extreme magnetosphere compresson to the moment immediately after the compression on 21 January 2005 (relations (A1) and (A2)), can be explained by a compression of the magnetic trap and adiabatical acceleration of RC particles.

**Acknowledgments.** The authors thank NASA/GSFC ISTP for providing data from the ACE, GOES, LANL and Cluster satellites. We are grateful to Cluster CIS and FGM instrument teams who have created the CIS (CODIF and HIA) and FGM instruments, and made their data available to the community. In particular, we greatly appreciate the anonymous Reviewer for providing the Cluster plasma data from the more recent calibrations of the CIS team. We thank C. W. Smith from the University of New Hampshire for providing the ACE magnetic data, and R. Skoug from Los Alamos National Laboratory for providing the ACE plasma data. We also thank NASA and NOAA for providing the GOES magnetic data, Los Alamos National Laboratory for providing the LANL plasma data, and Kyoto World Data Center for Geomagnetism for providing the Dst and ASY/SYM indices. Simulation results of SWMF/BATS-R-US code have been provided by the Community Coordinated Modeling Center at Goddard Space Flight Center through their public Runs on Request system (http://ccmc.gsfc.nasa.gov). The CCMC is a multi-agency partnership between NASA, AFMC, AFOSR, AFRL, AFWA, NOAA, NSF and ONR. This work was supported by grants NSC102-2111-M-008-023 from the National Science Council of Taiwan and by Ministry of Education under the Aim for Top University program 102G901-27 at National Central University of Taiwan as well as by RFBR grant 13-02-00461 and Programs P22, P26 of the RAS Presidium at MSU SINP and IKI.

## References

Bailey, J., and M. Gruntman (2013), Observations of exosphere variations during geomagnetic storms, *Geophys. Res. Lett.,40*, 1907-1911, doi:10.1002/grl.50443.

Baker, D. N., et al. (2013), A major solar eruptive event in July 2012: Defining extreme space weather scenarios, *Space Weather*, doi:10.1002/swe.20097.

Balogh, A., et al. (2001), The Cluster Magnetic Field Investigation: Overview of in-flight performance and initial results, *Ann. Geophys.*, *19*, 1207-1217.

Belov, A. V., et al. (2005), Ground level enhancement of the solar cosmic rays on January 20, 2005, *Proc. 29th Internat. Cosmic Ray Conf.* (Pune), *1*, 189-192.

Borovsky, J. E., M. F. Thomsen, and R. C. Elphic (1998), The driving of the plasma sheet by the solar wind, *J. Geophys. Res., 103*, 17,617–17,639, doi:10.1029/97JA02986.

Borrini, G., et al. (1982), Helium abundance enhancements in the solar wind, *J. Geophys. Res.*, *87(A9)*, 7370-7378, doi:10.1029/JA087iA09p07370.

Burlaga, L., et al. (1998), A magnetic cloud containing prominence material: January 1997, *J. Geophys. Res.*, *103(A1)*, 277-285, doi:0.1029/97JA02768.

Burton R. K., R. L. McPherron, and C. T. Russell (1975), An empirical relationship between interplanetary conditions and Dst, *J. Geophys. Res., 80*, 4204-4214, doi:10.1029/JA080i031p04204.

Cairns, I. H., et al. (1995), Unusual locations of Earth's bow shock on September 24-25, 1987: Mach number effects, *J. Geophys. Res., 100(A1)*, 47-62, doi:10.1029/94JA01978.

Chao, J.-K., et al. (2002), Models for the size and shape of the Earth's magnetopause and bow shock, in *Space Weather Study Using Multipoint Techniques*, Ed. L.-H. Lyu, 360 pp., Pergamon, New York, 127 – 135.

Chapman, S., and V. C. A. Ferraro (1931), A new theory of magnetic storms, *J. Geophys. Res.*, *36*, 77-97. *Nature*. 1930, 126, 129–130.

Claflin, E. S. (1970), Charge exchange cross section for hydrogen and helium ions incident on atomic hydrogen: 1 to 1000 keV, *Rep. TR-0059 (6260-20)-1. Aerosp. Corp.*, El Segundo, Calif., July 1970.

Cornwall, J. M., and M. Schulz (1979), Physics of heavy ions in the magnetosphere, *Solar System Plasma Physics*, Eds. E. N. Parker, C. F. Kennel, and L. J. Lanzerotti, North-Holland, Amsterdam, 3, 165-210.

Crooker, N. U., et al. (2000), Density extremes in the solar wind, *Geophys. Res. Lett., 27(23),* 3769–3772.

Chen, P. F. (2011), Coronal mass ejections: Models and their observational basis, *Living Rev. Solar Phys.*, *8*, 1–92.

Dandouras, I. S., et al. (2009), Magnetosphere response to the 2005 and 2006 extreme solar events as observed by the Cluster and Double Star spacecraft, *Adv. Space Res., 43*, 618-623, doi:10.1016/j.asr.2008.10.015.

DeForest, S. E. (1973), Detection of the solar wind at synchronous orbit, *J. Geophys. Res., 78,* 1195-1197.

Dmitriev, A., and A. Suvorova (2012), Equatorial trench at the magnetopause under saturation, *J. Geophys. Res.*, *117*, A08226, doi:10.1029/2012JA017834.

Dmitriev, A. V., J.-K. Chao, and D.-J. Wu (2003), Comparative study of bow shock models using Wind and Geotail observations, *J. Geophys. Res.*, *108(A12)*, 1464, doi:10.1029/2003JA010027.

Dmitriev, A. V., et al. (2005a), Geosynchronous magnetopause crossings on 29-31 October 2003, *J. Geophys. Res.*, *110*, A08209, doi:10.1029/2004JA010582.

Dmitriev, A. V., et al. (2005b), Indirect estimation of the solar wind conditions in 29–31 October 2003, *J. Geophys. Res.*, *110*, A09S02, doi:10.1029/2004JA010806.

Dmitriev, A., A. Suvorova, and J.-K. Chao (2011), A predictive model of geosynchronous magnetopause crossings, *J. Geophys. Res.*, *116*, A05208, doi:10.1029/2010JA016208.

Du, A. M., B. T. Tsurutani, and W. Sun (2008), Anomalous geomagnetic storm of 21–22 January 2005: A storm main phase during northward IMFs, *J. Geophys. Res.*, *113*, A10214, doi:10.1029/2008JA013284.

Fairfield, D. H. (1971), Average and unusual locations of the Earth's magnetopause and bow shock, *J. Geophys. Res.*, *76(28)*, 6700-6716, doi:10.1029/JA076i028p06700.

Feldstein, Ya. I., V. I. Pisarsky, N. M. Rudneva, and A. Grafe (1984) Ring current simulation in connection with interplanetary space conditions, *Planet. Space Sci.*, *32*, 975–984.

Fenrich, F. R., and J. G. Luhmann (1998), Geomagnetic response to magnetic clouds of different polarity, *Geophys. Res. Lett.*, *25*, 2999-3002.

Formisano, V. (1973), On the March 7-8, 1970 event, *J. Geophys. Res.*, *73*, 1198-1202.

Formisano, V. (1979), Orientation and shape of the Earth's bow shock in three dimensions, *Planet. Space Sci.*, *27*, 1151-1161.

Foullon, C., et al. (2007), Multi-Spacecraft Study of the 21 January 2005 ICME Evidence of Current Sheet Substructure Near the Periphery of a Strongly Expanding, Fast Magnetic Cloud, *Solar Phys.*, *244*, 139-165, doi:10.1007/s11207-007-0355-y.

Fritz, T. A., et al. (1974), Initial observations of magnetospheric boundaries by Explorer 45 ($S^3$), *Correlated Interplanetary and Magnetospheric Observations*, Ed. D. E. Page, Dordrecht–Holland: D. Reidel, pp. 485−506.

Ganushkina, N. Yu., M. W. Liemohn, M. V. Kubyshkina, R. Ilie, and H. J. Singer (2010), Distortions of the magnetic field by storm-time current systems in Earth's magnetosphere, *Ann. Geophys.*, 28, 123-140, doi:10.5194/angeo-28-123-2010.

Gopalswamy, N., et al. (2005), Introduction to violent Sun-Earth connection events of October - November 2003, *J. Geophys. Res.*, *110*, A09S00, doi:10.1029/2005JA011268.

Gosling, J. T., et al. (1980) Observations of large fluxes of $He^+$ in the solar wind following an interplanetary shock, *J. Geophys. Res.*, *85(A7)*, 3431-3434, doi: 10.1029/JA085iA07p03431.

Gosling, J. T., et al. (1982), Evidence for quasi-stationary reconnection at the dayside magnetopause, *J. Geophys. Res.*, *87(A4)*, 2147-2158, doi:10.1029/JA087iA04p02147.

Hoffman R. A., L. J. Cahill, Jr., R. R. Anderson, et al. (1975), Explorer 45($S^3$-A) observations of the magnetosphere and magnetopause during August 4-6, 1972, magnetic storm period, *J. Geophys. Res.*, *80(31)*, 4287-4296.

Hudson, M. K., et al. (1997), Simulations of radiation belt formation during storm sudden commencements, *J. Geophys. Res.*, *102(A7)*, 14087-14102, doi:10.1029/97JA03995.

Hudson, M. K., et al. (2004), 3D modeling of shock-induced trapping of solar energetic particles in the Earth's magnetosphere, *J. Atmosph. Solar-Terr. Phys.*, *66*, 1389-1397, doi:10.1016/j.jastp.2004.03.024.

Kistler, L. M., et al. (1989), Energy spectra of the major ion species in the ring current during geomagnetic storms, *J. Geophys. Res.*, *94(A4)*, 3579–3599.

Kovtyukh, A. S. (2001), Geocorona of hot plasma, *Cosmic Res.*, *39(6)*, 527-558.

Kovtyukh, A. S. (2010), Radial profile of pressure in a storm ring current as a function of $D_{st}$, *Cosmic Res.*, *48(3)*, 211–231.

Kuznetsov, S. N., and A. V. Suvorova (1998), An empirical model of the magnetopause for broad ranges of solar wind pressure and Bz IMF, in *Polar Cap Boundary Phenomena*, NATO ASI Ser., Eds. J. Moen, A. Egeland, and M. Lockwood, pp. 51-61, Kluwer Acad., Norwell, Mass.

Kuznetsov, S. N., et al. (2005), Proton acceleration during 20 January 2005 solar flare: CORONAS-F observations of high-energy gamma emission and GLE, 29th *International Cosmic Ray Conference, Pune, 1*, 49-52.

Kuznetsova, T. V., and A. I. Laptukhov (2011), Contribution of geometry of interaction between interplanetary and terrestrial magnetic fields into global magnetospheric state and geomagnetic activity, *Adv. Space Res., 47*, 978-990.

Lazutin, L. L., and S. N. Kuznetsov (2008), Nature of sudden auroral activations at the beginning of magnetic storms, *Geomag. and Aeron., 48(2)*, 165–174.

Lazutin, L. L., et al. (2010), Dynamics of solar protons in the Earth's magnetosphere during magnetic storms in November 2004 – January 2005, *Geomag. and Aeron., 50(2)*, 176-188.

Lockwood, J. A., L. Hsieh, and J. J. Quenby (1975), Some unusual features of the cosmic ray storm in August 1972, *J. Geophys. Res., 80,* 1725-1734, doi:10.1029/JA080i013p01725.

Lu, G., et al. (1998), Global energy deposition during the January 1997 magnetic cloud event, *J Geophys. Res., 103(A6),* 11,685-11,694, doi:10.1029/98JA00897.

McPherron, R. L., and T. P. O'Brien (2001), Predicting geomagnetic activity: The Dst index, in Space Weather, *Geophys. Monogr. Ser., vol. 125,* edited by P. Song, H. J. Singer, and G. L. Siscoe, 339-345, AGU, Washington, D. C.

McKenna-Lawlor, S., et al. (2010), Moderate geomagnetic storm (21-22 January 2005) triggered by an outstanding coronal mass ejection viewed via energetic neutral atoms, *J. Geophys. Res., 115,* A08213, doi:10.1029/2009JA014663.

Ober, D. M., M. F. Thomsen, and N. C. Maynard (2002), Observations of bow shock and magnetopause crossings from geosynchronous orbit on 31 March 2001, *J. Geophys. Res., 107(A8),* 1206, doi:10.1029/2001JA000284.

O'Brien, T. P., and R. L. McPherron (2000), An empirical phase space analysis of ring current dynamics: Solar wind control of injection and decay, *J. Geophys. Res.*, *105*, 7707- 7719.

O'Brien, T. P., and R. L. McPherron (2002), Seasonal and diurnal variation of Dst dynamics, *J. Geophys. Res., 107(A11),* 1341, doi:10.1029/2002JA009435.

Østgaard, N. (2003), Neutral hydrogen density profiles derived from geocoronal imaging, *J. Geophys. Res., 108(A7),* 1300, doi:10.1029/2002JA009749.

Peredo, M., et al. (1995), Three-dimensional position and shape of the bow shock and their variations with Alfvenic, sonic, and magnetosonic Mach numbers and interplanetary magnetic field orientation, *J. Geophys. Res.*, *100*, 7907-7916, doi:10.1029/94JA02545.

Pudovkin, M. I., S. A. Zaitseva, and I. Z. Sizova (1985), Growth rate and decay of magnetospheric ring current, *Planet. Space Sci., 33(10),* 1097-1102.

Rème, H., et al. (2001), First multispacecraft ion measurements in and near the Earth's magnetosphere with the identical Cluster Ion Spectrometry (CIS) experiment, *Ann. Geophys.*, *19,* 1303-1354.

Richard, R. L., et al. (2009), Modeling the entry and trapping of solar energetic particles in the magnetosphere during the November 24-25, 2001 storm, *J. Geophys. Res.*, *114*, A04210, doi:10.1029/2007JA012823.

Russell, C. T., and S. M. Petrinec (1996), Comment on "Towards an MHD theory for the standoff distance of Earth's bow shock" by I. H. Cairns and C. L. Grabbe, *Geophys. Res. Lett., 23,* 309-310, doi:10.1029/95GL03505.

Russell, C. T., et al. (2000), The extreme compression of the magnetosphere, May 4, 1998, as observed by the POLAR spacecraft, *Adv. Space Res.*, *25,* 1369-1375. doi:10.1016/S0273-1177(99)00646-8.

Russell, C. T., et al. (2013), The very unusual interplanetary coronal mass ejection of 2012 July 23: A blast wave mediated by solar energetic particles, *Astrophys. Jour.*, *770*, 38.

Schwenn R. (1983), Direct correlations between coronal transients and interplanetary disturbances, *Space Science Reviews*, *34(1),* 85-99.

Sharma, R., et al. (2013), Interplanetary and geomagnetic consequences of 5 January 2005 CMEs associated with eruptive filaments, *J. Geophys. Res.*, *118*, 3954-3967, doi:10.1002/jgra.50362.

Song, P., et al. (2001), Polar observations and model predictions during May 4, 1998, magnetopause, magnetosheath, and bow shock crossings, *J. Geophys. Res.*, *106*, 18,927-18,942, doi:10.1029/2000JA900126.

Spreiter, J. R., A. L. Summers, and A. Y. Alksne (1966), Hydromagnetic flow around the magnetosphere, *Planet. Space Sci.*, *14*, 223-253.

Suvorova A. V, A. V. Dmitriev, S. N. Kuznetsov (1999), Dayside magnetopause models, *Rad. Meas.*, *30(5)*, 687-692.

Suvorova A., et al. (2005), Necessary conditions for the geosynchronous magnetopause crossings, *J. Geophys. Res.*, *110*, A01206, doi:10.1029/2003JA010079.

Troshichev, O., D. Sormakov, A. Janzhura (2011a), Relation of PC index to the geomagnetic storm Dst variation, *J. Atmosph. Solar-Terr. Phys.*, *73*, 611-622.

Troshichev, O. A., N. A. Podorozhkina, A. S. Janzhura (2011b), Relationship between PC index and magnetospheric substorms observed under conditions of northward IMF, *J. Atmosph. Solar-Terr. Phys.*, *73*, 2373-2378.

Tsurutani, B. T., et al. (2003), The extreme magnetic storm of 1-2 September 1859, *J. Geophys. Res.*, *108(A7)*, 1268, doi:10/1029/2002JA009504.

Tsyganenko, N. A. (2000), Solar wind control of the tail lobe magnetic field as deduced from Geotail, AMPTE/IRM, and ISEE 2 data, *J. Geophys. Res.*, *105*, 5517–5528, doi:10.1029/1999JA000375.

Tsyganenko, N. A., and M. I. Sitnov (2005), Modeling the dynamics of the inner magnetosphere during strong geomagnetic storms, *J. Geophys. Res.*, *110*, A03208, doi:10.1029/2004JA010798.

Vaisberg, O., and G. Zastenker (1976), Solar wind and magnetosheath observations at earth during August 1972, *Space Science Reviews*, *19*, 687-702.

Verigin, M. I., et al. (2001), Analysis of the 3-D shape of the terrestrial bow shock by Interball/MAGION 4 observations, *Adv. Space Res.*, *28*, 857-862.

Wang, C. B., J.-K. Chao, and C. H. Lin (2003), Influence of the solar wind dynamic pressure on the decay and injection of the ring current, *J. Geophys. Res.*, *108(A9)*, 1341, doi:10.1029/2003JA009851.

**Table 1.** Observations of the bow shock ($R_{BS}$) and magnetopause ($R_{MP}$) inside geosynchronous orbit

| Date<br>time | Satellites | $R_{BS}$, Re<br>(zenith<br>angle)<br>duration | $R_{MP}$, Re<br>(zenith<br>angle) | *V*,<br>km/s | ***D***<br>$cm^{-3}$ | *Pd*<br>nPa | IMF B<br>Bz<br>nT | Ph* | Reference |
|---|---|---|---|---|---|---|---|---|---|
| 1970<br>8 Mar<br>~20 UT | ATS-5<br>HEOS-1 | ≤6.6(0°)<br>3 min | <6.6(0°) | 880 | ~40 | >50 | ~30<br>>0 | 0 | *DeForest, 1973*<br>*Formisano,* 1973 |
| 1972<br>4 Aug<br>~23 UT | Explorer-45<br>Prognoz-2<br>HEOS-2 | ~10(75°)<br>- | 5.2(45°)<br>6(40°) | 1700 | ~30 | ≥150 | >50<br>± | 2 | *Hoffman et al., 1975*<br>*Lockwood et al.,*1975<br>*Vaisberg and Zastenker,*1976 |
| 1998<br>4 May<br>~07 UT | Polar<br>WIND<br>ACE | 7.3(32°)<br>2 min | 5.3(19°) | 800 | 60 | >65 | 20<br>>0 | 2 | *Russell et al., 2000*<br>*Song et al.,* 2001 |
| 2001<br>31 Mar<br>~05 UT | 1994-084<br>ACE<br>IMP8 | ≤6.6(0°)<br>10 min | <6.6(90°) | 700 | ~70 | >60 | ~50<br><0 | 0 | *Ober et al., 2002* |
| 2003<br>30 Oct<br>22UT | GOES-10<br>ACE | ≤6.6(15°)<br>2 min | <6.6 | 1200 | - | >40 | 20<br>>0 | 2 | *Dmitriev et al.,* 2005a,b<br>*Veselovskyi et al.,* 2004 |
| 2005<br>21 Jan<br>~19UT | GOES<br>DBST-1<br>ACE<br>Cluster | ≤6.6(0°)<br>40 min<br>2 hours** | <6.6 | 1000 | ~60 | >90 | 40<br>± | 1 | *Foullon et al.,* 2007<br>*Du et al.,* 2008<br>*Dandouras et al.,* 2009<br>*McKenna-Lawlor et al., 2010* |

* Phase of the storm: 0 - onset, 1 - main, 2 - recovery
** Shown at present study

**Table 2.** Predicted bow shock subsolar distances and dynamic pressure

| Model | *R*s, Re | *Pd*, nPa |
|---|---|---|
| *Formisano* [1979] | 4.4 | 1500 |
| *Cairns et al.* [1995] | 4.5 | - |
| *Peredo et al.* [1995] | 5.3 | 1500 |
| *Russell and Petrinec* [1996] | 5.4 | 140 |
| BSV [*Verigin et al.*, 2001] | 5.2 | 80 |
| BSC [*Chao et al.*, 2002] | 8. | 75 |

## Figure Captions

**Figure 1.** Heliospheric and geomagnetic conditions during magnetic storm on 21 January 2005 (from top to bottom): fluxes of solar energetic particles (SEP), solar wind velocity *V*, solar wind proton density *D*, helium contribution; solar wind dynamic pressure *P*d, IMF Bx, By and Bz components in GSM, *AE* and *Dst* geomagnetic indices. Solar wind plasma and IMF parameters measured by ACE and Cluster are shown, respectively, by red and blue curves. The time profiles of ACE and Cluster are shifted by the time of the solar wind propagation. Hourly *Dst* and 1-min SYM-H index are shown by gray histogram and black curves. At the bottom panel, *Dst** is corrected by pressure acquired from ACE (red curve) and Cluster (blue curve). See details in the text.

**Figure 2.** Prediction results for the 1-hour Dst variation during magnetic storm on 21 – 22 January 2005 using different empirical models. The beginning of the storm from 17 to 19 UT is predicted quite well. The models fail after 19 UT, when the Dst index continues decreasing despite of northward IMF orientation.

**Figure 3.** Location in GSM of geosynchronous and high-apogee satellites at ~1850 UT on 21 January 2005 in X-Y plane (left panel) and X-Z plane (right panel). In the X-Y plane, the position of bow shock (red curve) and magnetopause (blue curve) are calculated, respectively, by BSV [*Verigin et al.*, 2001] and KS98 [*Suvorova* et al., 1999] empirical models for the extreme solar wind conditions. Under such conditions, the subsolar bow shock and whole dayside magnetopause are located inside geosynchronous orbit.

**Figure 4.** Geosynchronous magnetopause (vertical blue dashed lines) and bow shock (vertical red dashed lines) crossings observed by GOES-10 (a) and GOES-12 (b) on 21 January 2005. The panels show (from top to bottom): solar wind dynamic pressure calculated from the ACE (red curve) and Cluster (blue curve) data and modeled dynamic pressure *P*gmc required for magnetopause geosynchronous crossing [*Dmitriev et al.*, 2011]; geocentric distance to the bow shock modeled by BSC model [*Chao et al.*, 2002] for the ACE (red curve) and Cluster (blue curve) pressure, and geocentric distance to the magnetopause (black curve) modeled by KS98 model [*Suvorova et al.*, 1999] for the Cluster pressure; *B*z, *B*y and *B*x observed by the satellites GOES (black curves) and Cluster-3 (blue curves) and predicted by a global MHD model (red curves); magnetic local time of GOES. At panel (b), the bow shock distance was calculated for the Cluster pressure by the BSC model (red) and by a model BSV [*Verigin et al.*, 2001]. The magnetopause and bow shock were calculated for the corresponding GOES angular location. Note that GOES-10 and GOES-12 were situated in the interplanetary medium from 1845 to 2035 UT and from 1845 to 2010 UT, respectively.

**Figure 5.** Geosynchronous magnetopause crossing (vertical blue dashed line) observed by LANL-1997 on 21 January 2005. The panels show (from top to bottom): solar wind dynamic pressure calculated from Cluster (blue curve) data and modeled dynamic pressure *P*gmc; KS98 model prediction of the geocentric distance to the magnetopause (black curve) for the Cluster pressure; plasma ratios RI (red curve) and RE (blue curve, see details in the text); magnetic local time. LANL-1997 encountered with the magnetosheath at ~0530MLT.

**Figure 6.** Temporal dynamics of pitch-angle anisotropy for the protons with energies >30 keV (black crosses) and >100 keV (blue triangles) observed by POES-17 near the noon-midnight meridian at $L$ ~ 5 on 21 January 2005. From ~19 to ~22 UT (restricted by red dashed lines), the anisotropy was less than or about 1 indicating that the majority of protons were not trapped at $L$ ~ 5.

**Figure 7.** Dynamics of proton integral spectra observed by POES-17 satellite near the noon-midnight meridian on 21 – 22 January 2005: (a) at $L = 2$; (b) at $L = 3$; (c) at $L = 4$; (d) at $L = 5$. Different

symbols and colors correspond to different observation times: blue diamonds – 17 UT, red triangles – 21 UT on 21 January and green diamonds – 05 UT on 22 January. At 21 UT, the fluxes of low-energy protons (<1 MeV) increased in the inner magnetosphere ($L < 4$) by more than 10 times.

**Figure 8.** Temporal variations of 1 – 5 MeV protons observed by CORONAS-F satellite on 21 – 22 January, 2005. Black curve with squares corresponds to a region of $L = 1 – 2$, red curve with triangles – $L = 2 – 3$, blue curve with diamonds – $L = 3 – 4$, and pink curve with crosses – $L = 4 – 5$. After 18UT on 21 January, the proton fluxes increased substantially in the inner magnetosphere.

**Figure 9.** Observed and proposed variations of the solar wind plasma and geomagnetic parameters on 21 January 2005 (from top to bottom): helium contribution He/H measured by Cluster C4 (blue curve) and 4-time magnified one (red curve); solar wind dynamic pressure *P*d calculated from Cluster C-4 data (blue curve) and with using the magnified He/H (red curve); nose distances to the bow shock and magnetopause predicted by the models BSC(red curve), BSV (blue curve) and KS98 (black curve) for the magnified He/H; *Dst* variation observed (black curve) and normalized by the observed *P*d (blue curve) and by the magnified *P*d (red curve) as well as a driving parameter $\varepsilon$ for the tail current (dotted green curve, right axis). The vertical red dashed lines restrict the interplanetary interval when the subsolar magnetopause was located upstream of the bow shock. The assumption of strong helium contribution of ~30% allows resolving the discrepancies between the observations and model predictions.

**Figure 10.** Variations of plasma parameters measured by Cluster on 21 January 2005 (from top to bottom): CODIF C4 energy-time spectrograms (in particle energy flux units) for H+ and He++; the corresponding densities of H+ and He++; Cluster C1 HIA (no mass discrimination) ion energy-time spectrogram and corresponding density. The data come from the more recent calibrations of the CIS team (acquired from private communication with anonymous Reviewer of this paper).

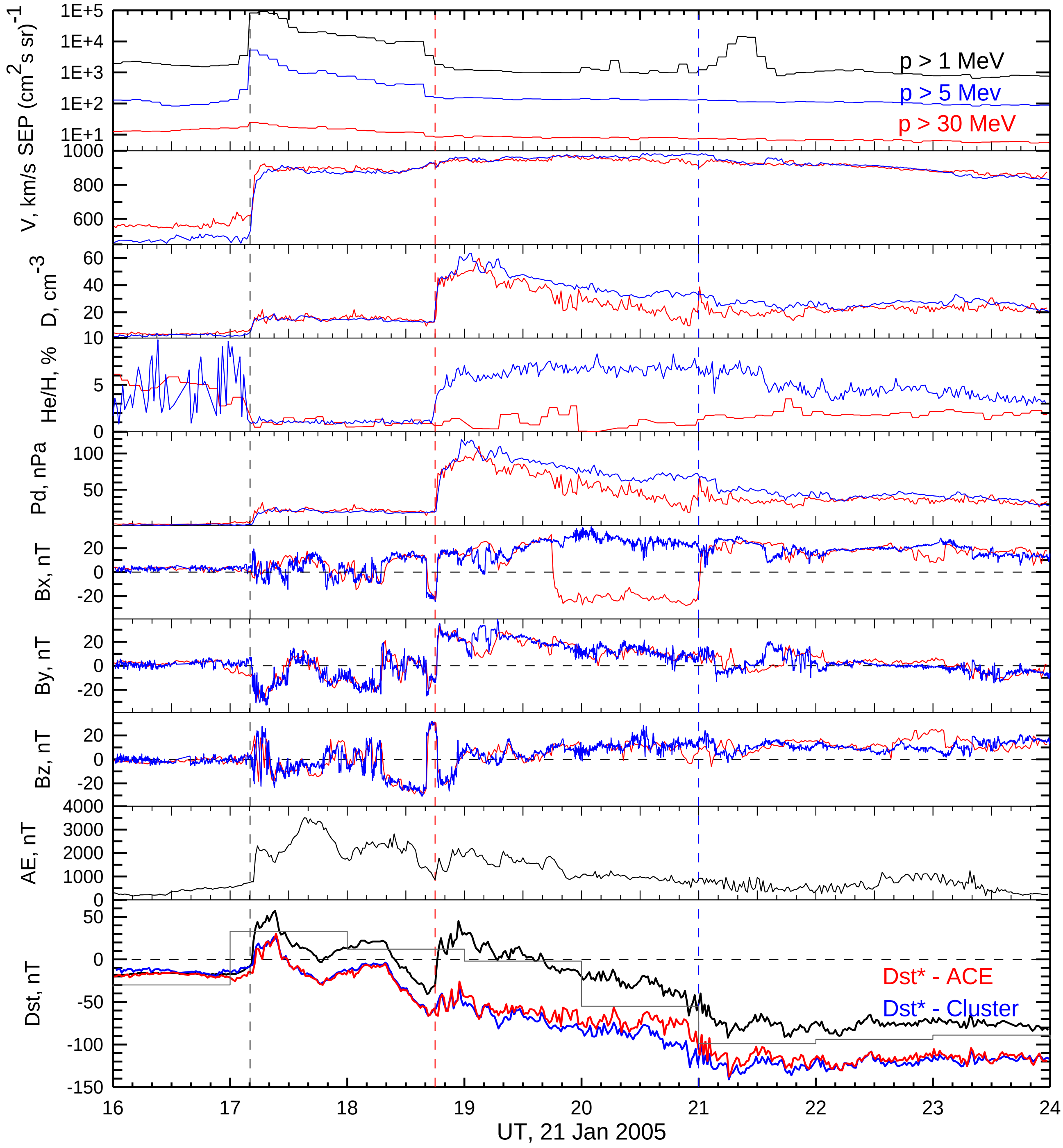

Figure 1. Heliospheric and geomagnetic conditions during magnetic storm on 21 January 2005 (from top to bottom): fluxes of solar energetic particles (SEP), solar wind velocity *V*, solar wind proton density *D*, helium contribution; solar wind dynamic pressure *P*d, IMF Bx, By and Bz components in GSM, *AE* and *Dst* geomagnetic indices. Solar wind plasma and IMF parameters measured by ACE and Cluster are shown, respectively, by red and blue curves. The time profiles of ACE and Cluster are shifted by the time of the solar wind propagation. Hourly *Dst* and 1-min SYM-H index are shown by gray histogram and black curves. At the bottom panel, *Dst** is corrected by pressure acquired from ACE (red curve) and Cluster (blue curve). See details in the text.

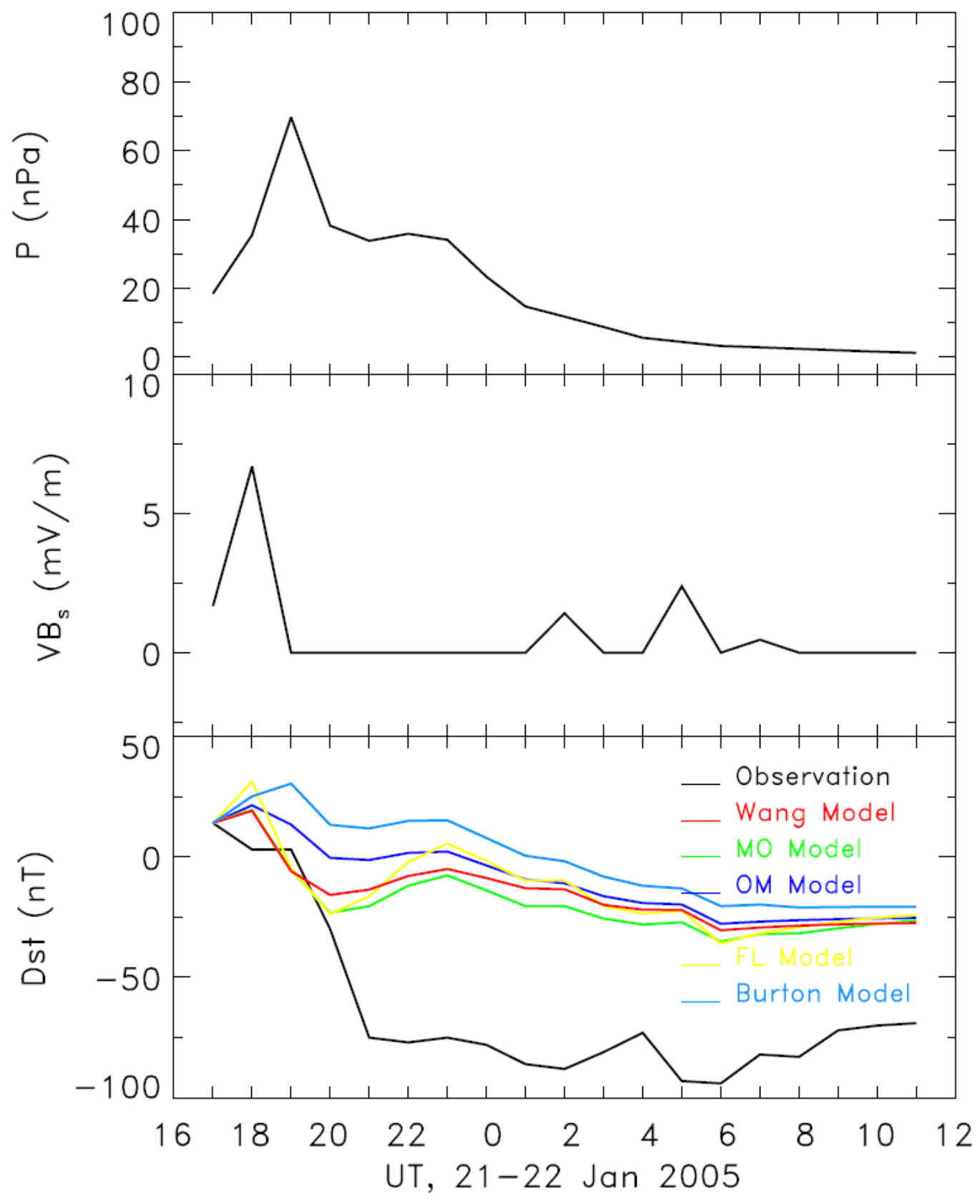

Figure 2. Prediction results for the 1-hour Dst variation during magnetic storm on 21 – 22 January 2005 using different empirical models. The beginning of the storm from 17 to 19 UT is predicted quite well. The models fail after 19 UT, when the Dst index continues decreasing despite of northward IMF orientation.

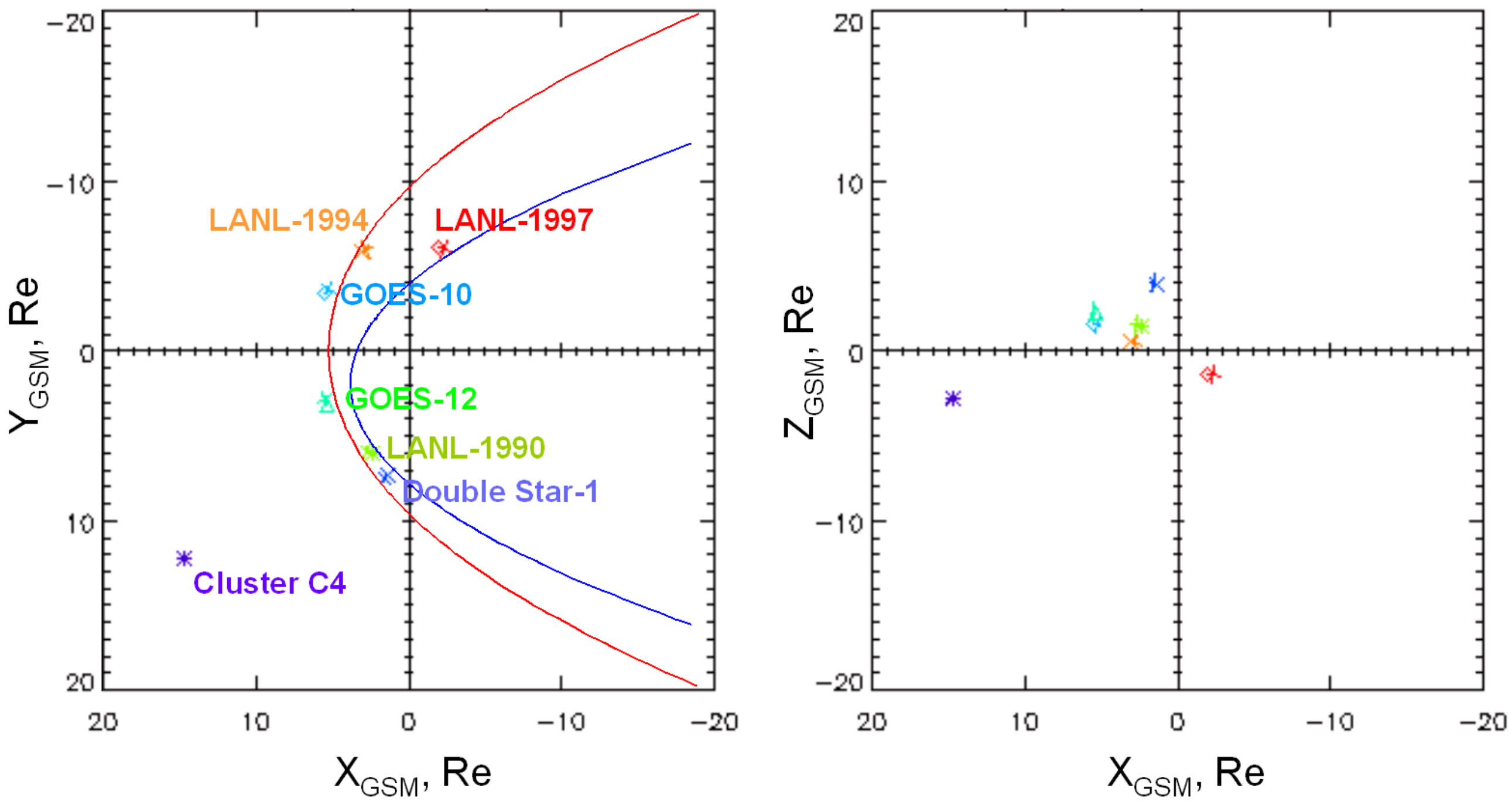

Figure 3. Location in GSM of geosynchronous and high-apogee satellites at ~1850 UT on 21 January 2005 in X-Y plane (left panel) and X-Z plane (right panel). In the X-Y plane, the position of bow shock (red curve) and magnetopause (blue curve) are calculated, respectively, by BSV [*Verigin et al.*, 2001] and KS98 [*Suvorova* et al., 1999] empirical models for the extreme solar wind conditions. Under such conditions, the subsolar bow shock and whole dayside magnetopause are located inside geosynchronous orbit.

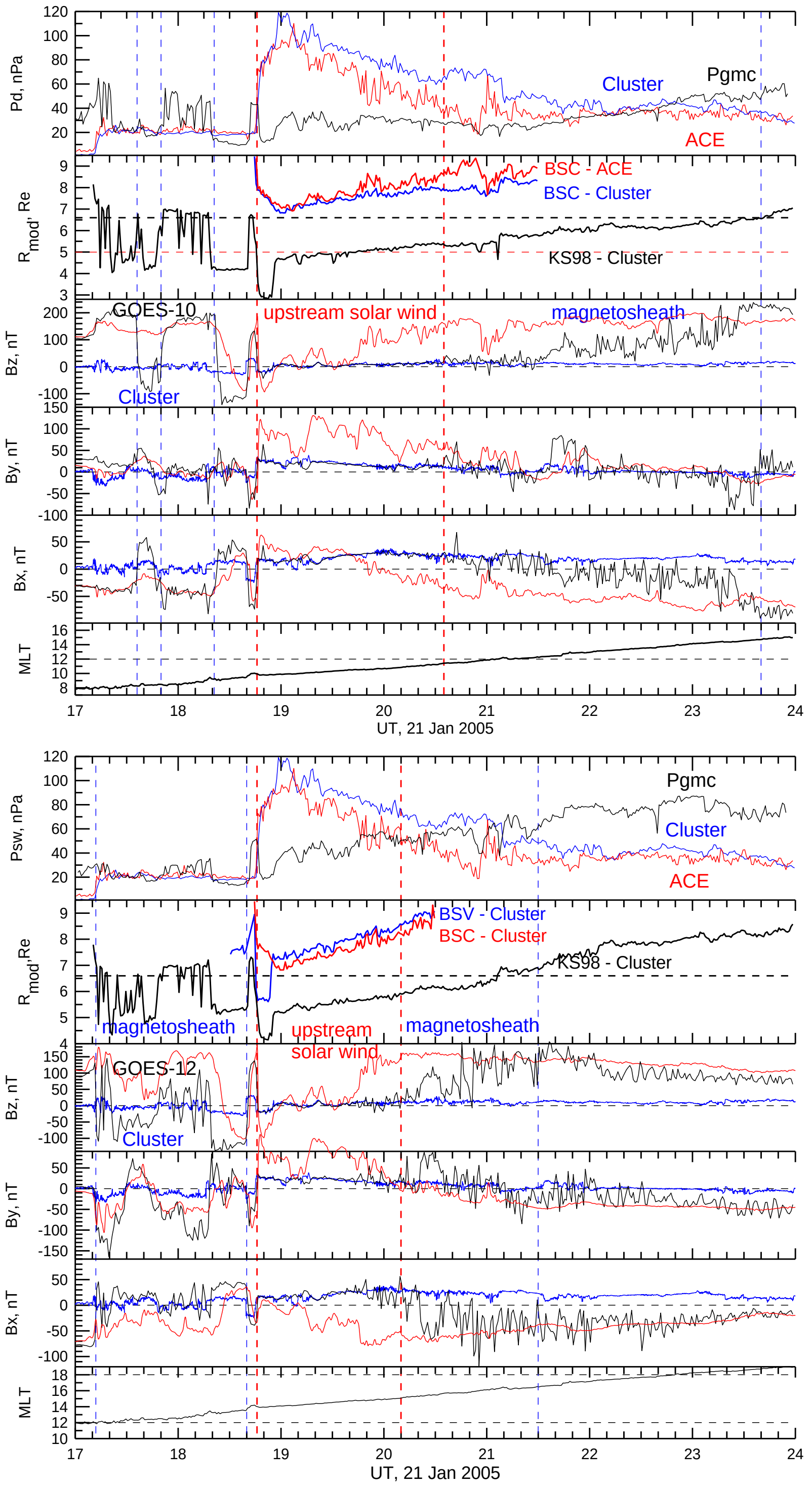

a

b

Figure 4. Geosynchronous magnetopause (vertical blue dashed lines) and bow shock (vertical red dashed lines) crossings observed by GOES-10 (a) and GOES-12 (b) on 21 January 2005. The panels show (from top to bottom): solar wind dynamic pressure calculated from the ACE (red curve) and Cluster (blue curve) data and modeled dynamic pressure *P*gmc required for magnetopause geosynchronous crossing [*Dmitriev et al.*, 2011]; geocentric distance to the bow shock modeled by BSC model [*Chao et al.*, 2002] for the ACE (red curve) and Cluster (blue curve) pressure, and geocentric distance to the magnetopause (black curve) modeled by KS98 model [*Suvorova et al.*, 1999] for the Cluster pressure; *Bz*, *By* and *Bx* observed by the satellites GOES (black curves) and Cluster-3 (blue curves) and predicted by a global MHD model (red curves); magnetic local time of GOES. At panel (b), the bow shock distance was calculated for the Cluster pressure by the BSC model (red) and by a model BSV [*Verigin et al.*, 2001]. The magnetopause and bow shock were calculated for the corresponding GOES angular location. Note that GOES-10 and GOES-12 were situated in the interplanetary medium from 1845 to 2035 UT and from 1845 to 2010 UT, respectively.

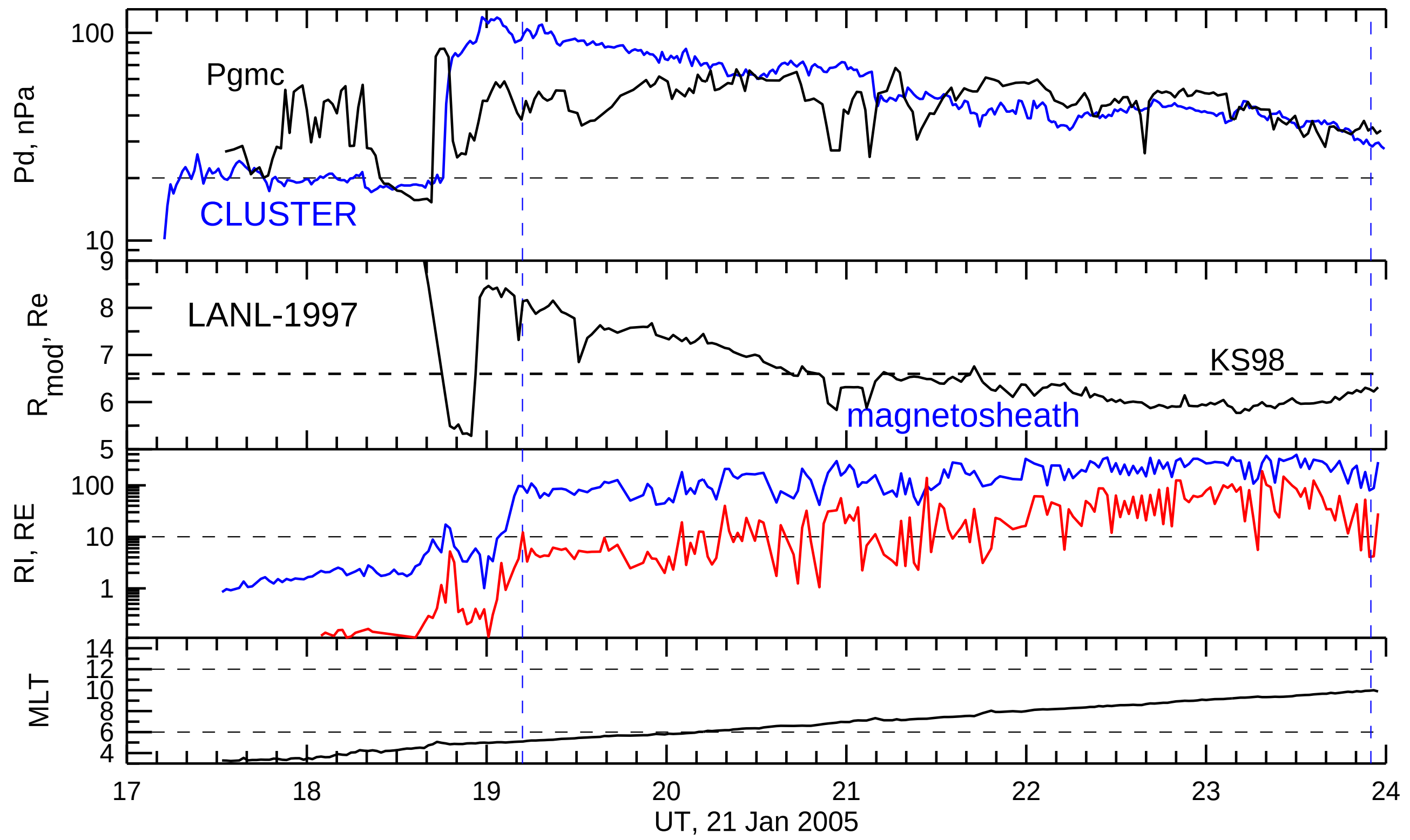

Figure 5. Geosynchronous magnetopause crossing (vertical blue dashed line) observed by LANL-1997 on 21 January 2005. The panels show (from top to bottom): solar wind dynamic pressure calculated from Cluster (blue curve) data and modeled dynamic pressure *P*gmc; KS98 model prediction of the geocentric distance to the magnetopause (black curve) for the Cluster pressure; plasma ratios RI (red curve) and RE (blue curve, see details in the text); magnetic local time. LANL-1997 encountered with the magnetosheath at ~0530MLT.

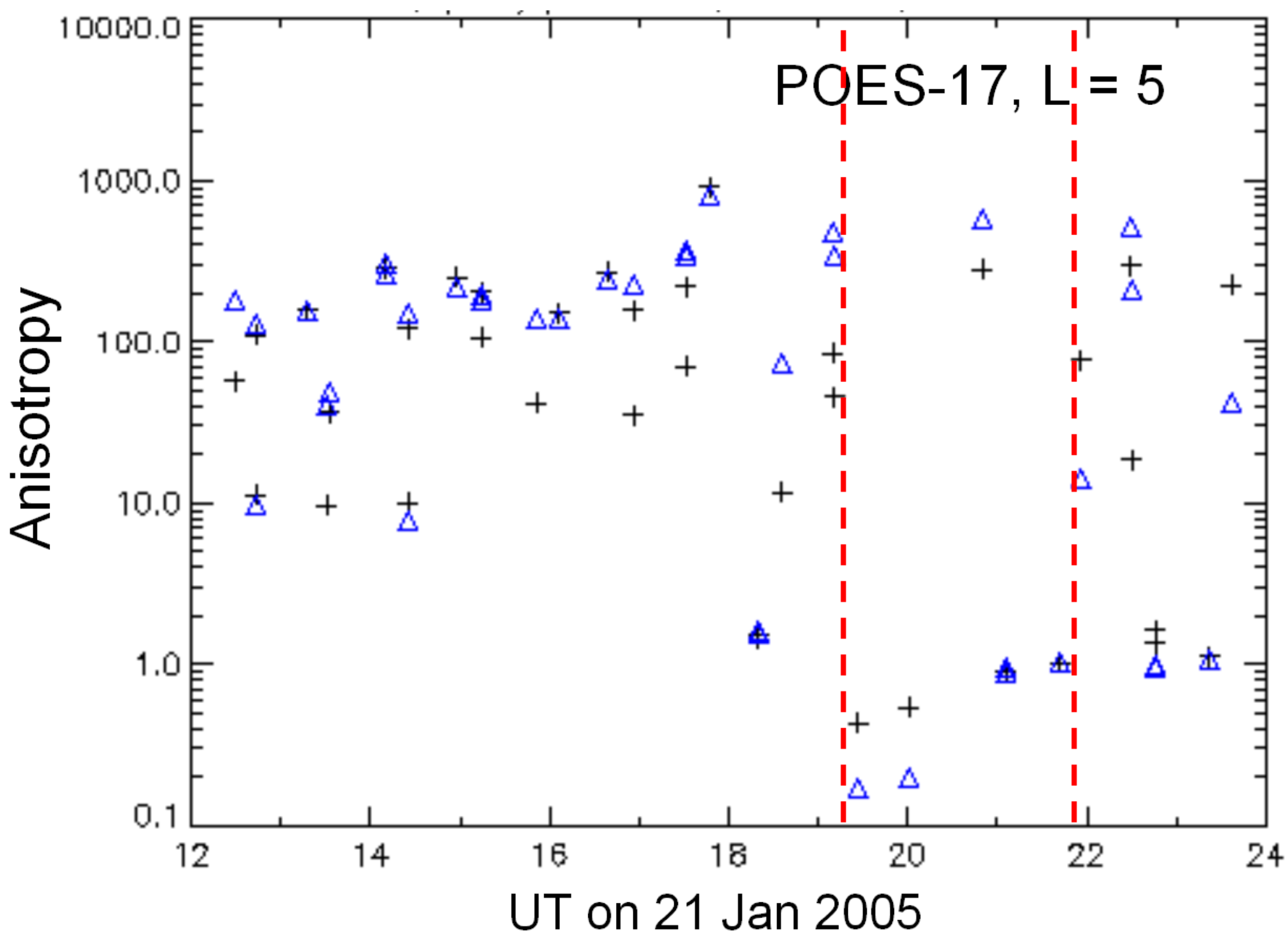

Figure 6. Temporal dynamics of pitch-angle anisotropy for the protons with energies >30 keV (black crosses) and >100 keV (blue triangles) observed by POES-17 near the noon-midnight meridian at $L$ ~ 5 on 21 January 2005. From ~19 to ~22 UT (restricted by red dashed lines), the anisotropy was less than or about 1 indicating that the majority of protons were not trapped at $L$ ~ 5.

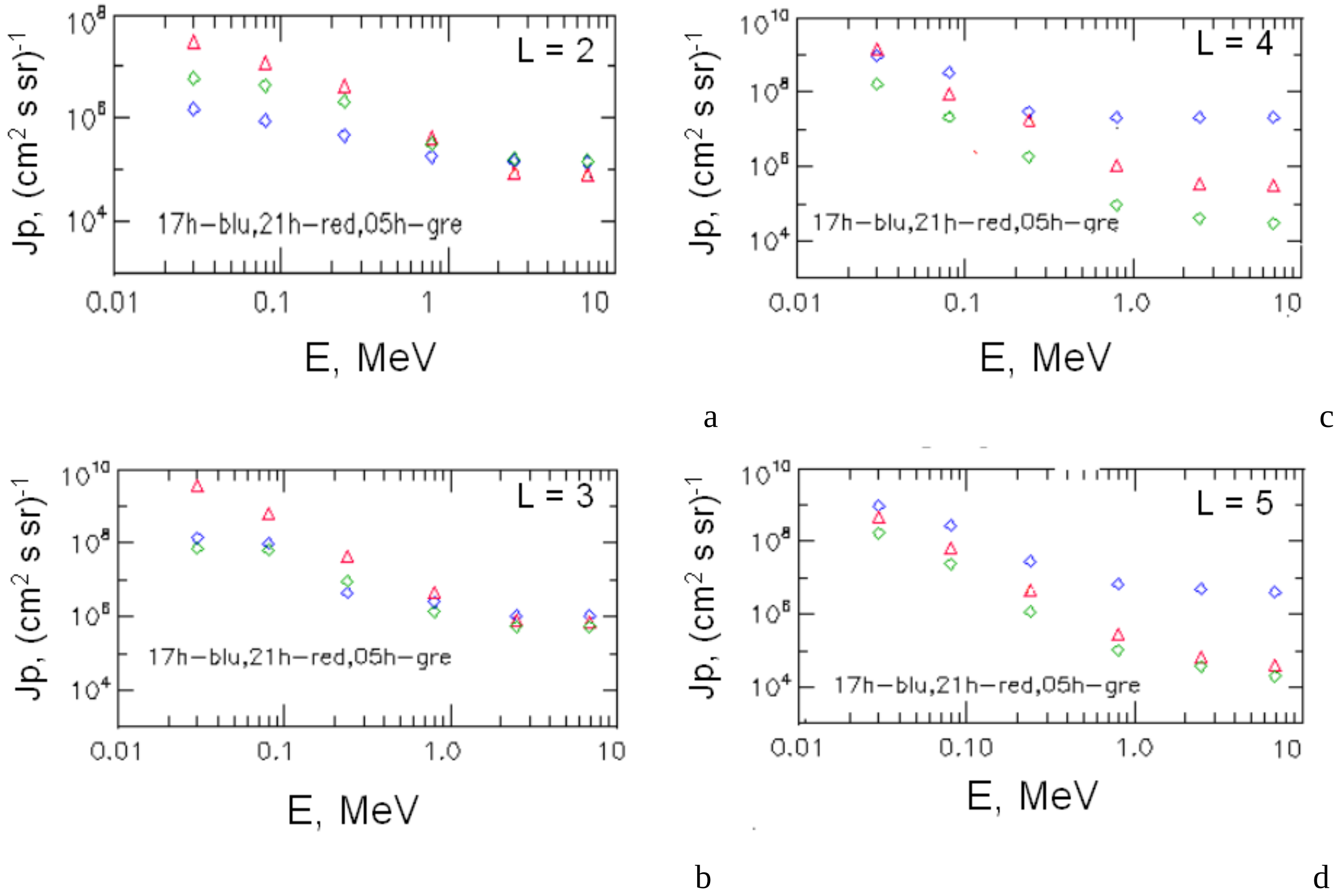

Figure 7. Dynamics of proton integral spectra observed by POES-17 satellite near the noon-midnight meridian on 21 – 22 January 2005: (a) at $L$ = 2; (b) at $L$ = 3; (c) at $L$ = 4; (d) at $L$ = 5. Different symbols and colors correspond to different observation times: blue diamonds – 17 UT, red triangles – 21 UT on 21 January and green diamonds – 05 UT on 22 January. At 21 UT, the fluxes of low-energy protons (<1 MeV) increased in the inner magnetosphere ($L$ < 4) by more than 10 times.

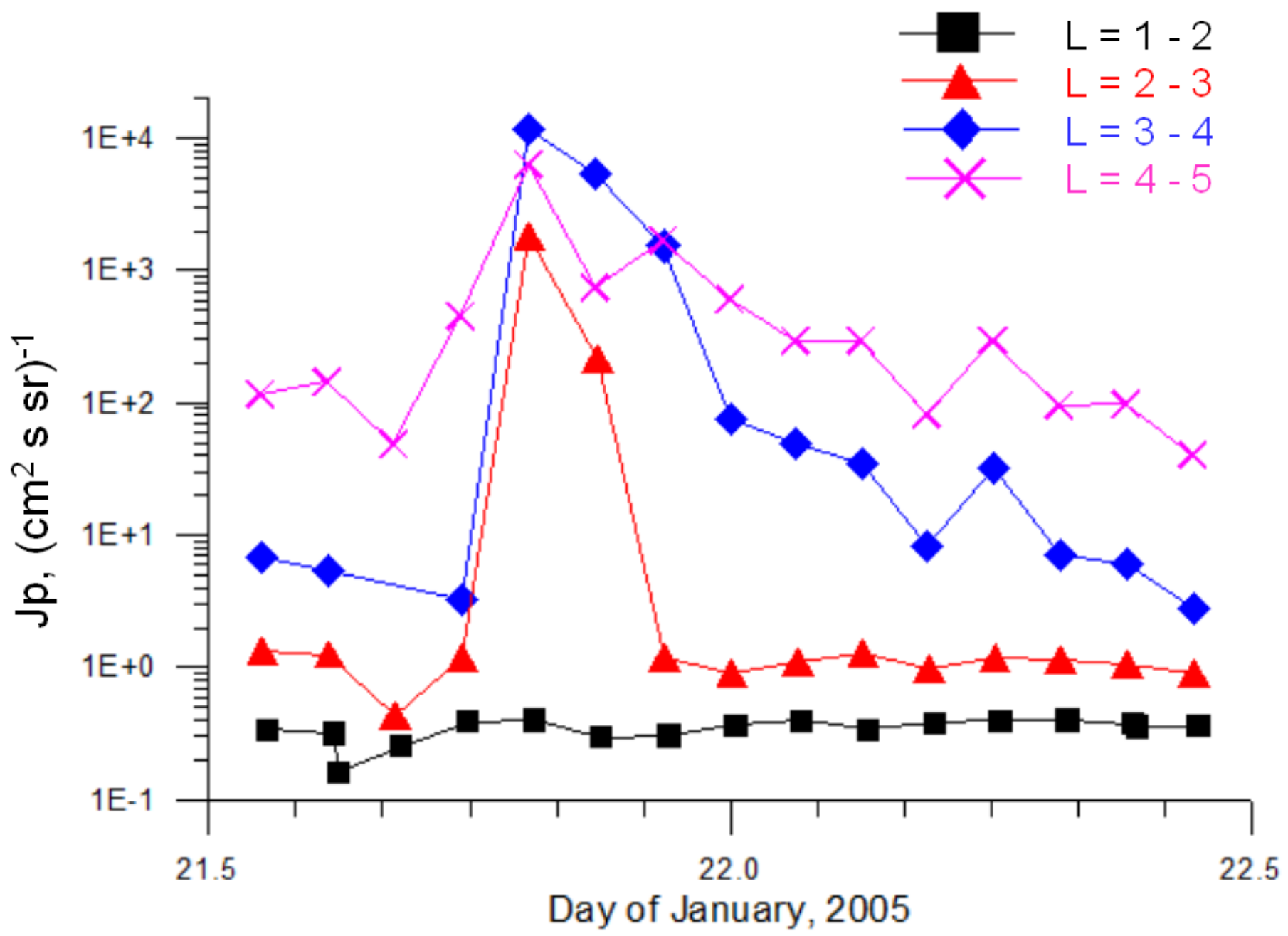

Figure 8. Temporal variations of 1 – 5 MeV protons observed by CORONAS-F satellite on 21 – 22 January, 2005. Black curve with squares corresponds to a region of $L$ = 1 – 2, red curve with triangles – $L$ = 2 – 3, blue curve with diamonds – $L$ = 3 – 4, and pink curve with crosses – $L$ = 4 – 5. After 18UT on 21 January, the proton fluxes increased substantially in the inner magnetosphere.

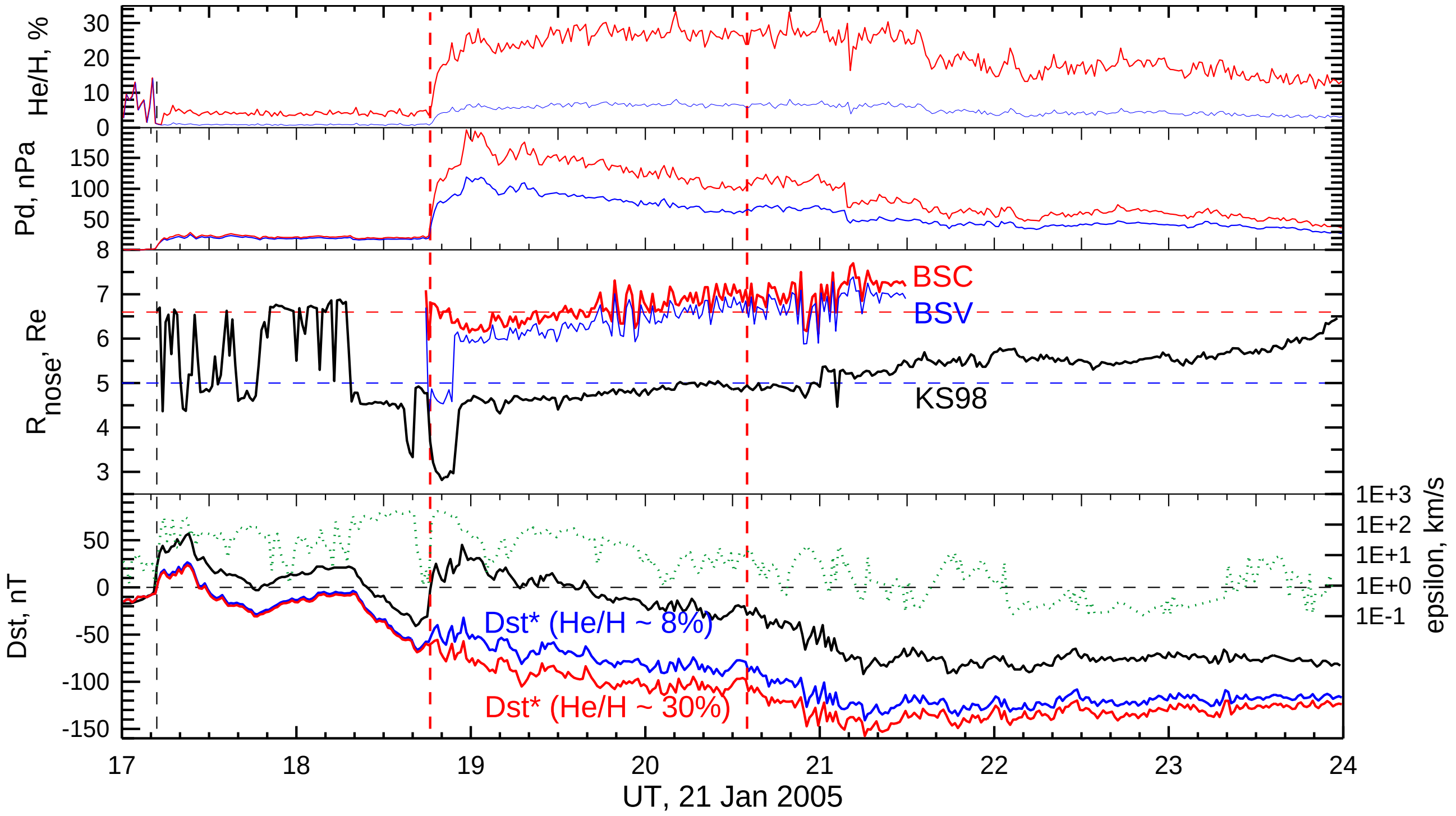

Figure 9. Observed and proposed variations of the solar wind plasma and geomagnetic parameters on 21 January 2005 (from top to bottom): helium contribution He/H measured by Cluster C4 (blue curve) and 4-time magnified one (red curve); solar wind dynamic pressure *P*d calculated from Cluster C-4 data (blue curve) and with using the magnified He/H (red curve); nose distances to the bow shock and magnetopause predicted by the models BSC(red curve), BSV (blue curve) and KS98 (black curve) for the magnified He/H; *Dst* variation observed (black curve) and normalized by the observed *P*d (blue curve) and by the magnified *P*d (red curve) as well as a driving parameter $\varepsilon$ for the tail current (dotted green curve, right axis). The vertical red dashed lines restrict the interplanetary interval when the subsolar magnetopause was located upstream of the bow shock. The assumption of strong helium contribution of ~30% allows resolving the discrepancies between the observations and model predictions.

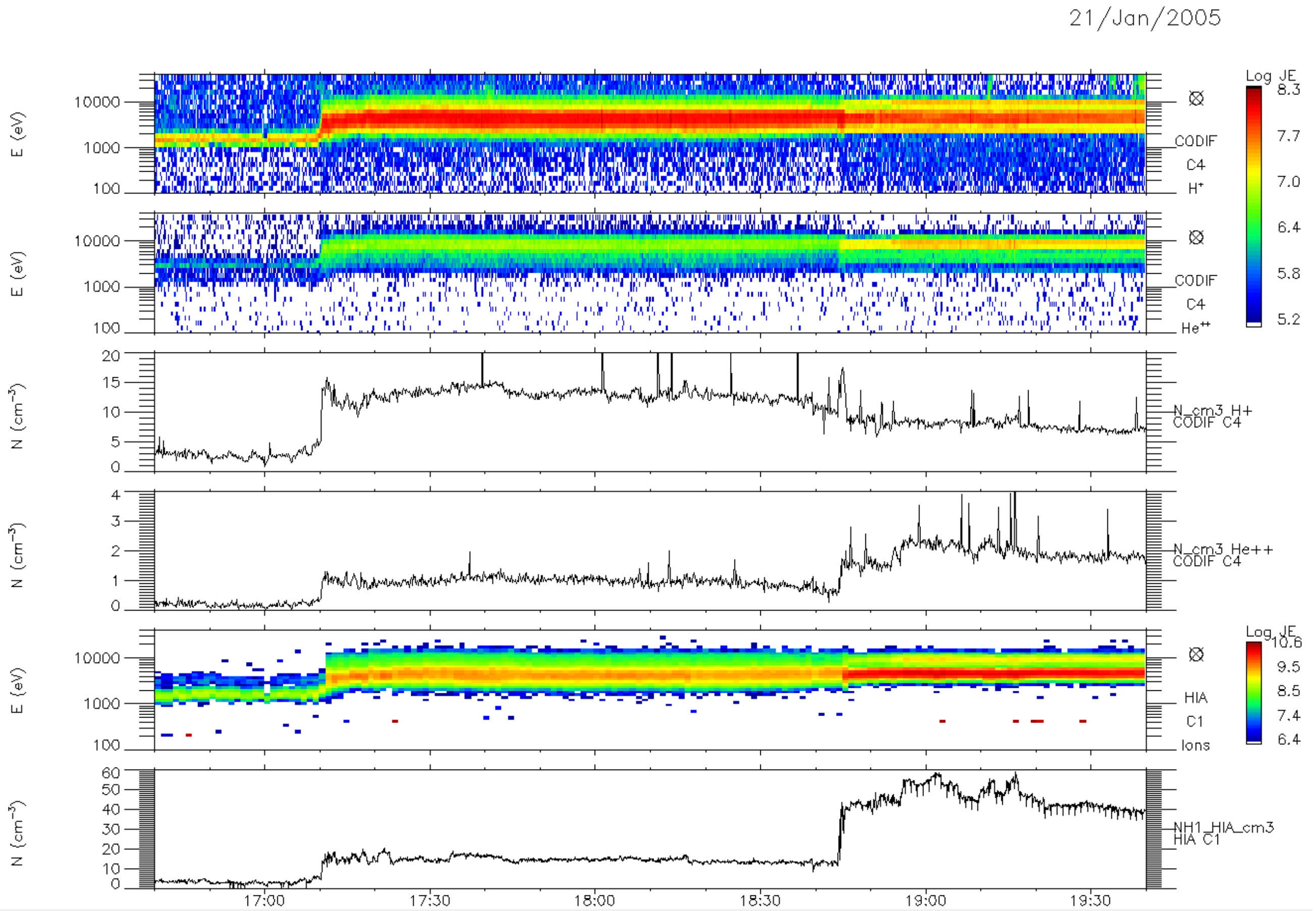

Figure 10. Variations of plasma parameters measured by Cluster on 21 January 2005 (from top to bottom): CODIF C4 energy-time spectrograms (in particle energy flux units) for H+ and He++; the corresponding densities of H+ and He++; Cluster C1 HIA (no mass discrimination) ion energy-time spectrogram and corresponding density. The data come from the more recent calibrations of the CIS team (acquired from private communication with anonymous Reviewer of this paper).